# Statistical PT-symmetric lasing in an optical fiber network

Ali K. Jahromi, Absar U. Hassan, Demetrios N. Christodoulides, and Ayman F. Abouraddy

*CREOL, The College of Optics & Photonics, University of Central Florida, Orlando, FL 32816, USA*

**PT-symmetry in optics is a condition whereby the real and imaginary parts of the refractive index across a photonic structure are deliberately balanced. This balance can lead to a host of novel optical phenomena, such as unidirectional invisibility, loss-induced lasing, single-mode lasing from multimode resonators, and non-reciprocal effects in conjunction with nonlinearities. Because PT-symmetry has been thought of as fragile, experimental realizations to date have been usually restricted to on-chip micro-devices. Here, we demonstrate that certain features of PT-symmetry are sufficiently robust to survive the statistical fluctuations associated with a macroscopic optical cavity. We construct optical-fiber-based coupled-cavities in excess of a kilometer in length (the free spectral range is less than 0.8 fm) with balanced gain and loss in two sub-cavities and examine the lasing dynamics. In such a macroscopic system, fluctuations can lead to a cavity-detuning exceeding the free spectral range. Nevertheless, by varying the gain-loss contrast, we observe that both the lasing threshold and the growth of the laser power follow the predicted behavior of a stable PT-symmetric structure. Furthermore, a statistical symmetry-breaking point is observed upon varying the cavity loss. These findings indicate that PT-symmetry is a more robust optical phenomenon than previously expected, and points to potential applications in optical fiber networks and fiber lasers.**

Since their mathematical inception [1,2], non-Hermitian PT-symmetric notions have found manifestations in many diverse physical embodiments, ranging from photonics [3-9] to acoustics [10,11], phononics [12], and even electronics [13,14]. Nevertheless, optics has proven to date to be the most convenient platform for the realization of PT-symmetry. In large part, the suitability of optics is a consequence of the facile deviation from Hermiticity achieved by adding optical loss (attenuation) or gain (amplification) to an initially unitary (lossless) system. These investigations have led to the observation of a number of counter-intuitive effects such as loss-induced transparency [15], lasing suppression with increased gain [16], and the revival of lasing with increased loss [17]. Furthermore, phenomena such as double refraction, power oscillations, and solitons have been observed in photonic lattices [18,19]. Recently, perfect absorption has been suggested [20] and realized [21] in PT-symmetric configurations, in addition to a burgeoning effort on PT-metasurfaces [22-27]. Furthermore, PT-symmetry has been employed to achieve one-sided invisibility [28,29], demonstrate unidirectional scattering and absorption [30,31], construct mode-selective lasers [32,33], and realize on-chip unidirectional devices [34,35]. Moreover, ramifications of these concepts are currently being extended to the emerging field of topological photonics [36-38] and ultra-sensitive measurement devices [39-41].

Exact PT-symmetry is achieved by arranging a delicate balance deliberately introduced into the spatial distribution of the complex refractive index throughout the system. In general, it is required to simultaneously maintain the real part of the refractive index *symmetric* while the imaginary part (corresponding to gain or loss) is kept *anti-symmetric* under inversion [3,4]. However, it has been shown that any patterned gain-loss structure in deterministic scheme may exhibits counter-intuitive PT-symmetric-related phenomena [42]. Because establishing a PT-symmetric refractive index distribution imposes stringent fabrication requirements, experimental realizations to date have been usually restricted to on-chip micro-devices. To date, optical demonstrations of PT-symmetry have focused on micro-structures realized on a chip, ranging from coupled semiconductor [15] or photorefractive waveguides [5] to coupled ring resonators in InP [32,43], quantum cascade disk-lasers [16], or erbium-doped silica [34,35]. The large free spectral range (FSR) associated with micro-devices helps isolate a single or a few relevant resonant modes in a compact and stable manner, thereby justifying a fully deterministic theoretical treatment. However, in larger optical systems, such as fiber networks, the FSR can be so small that unavoidable fluctuations lead to detuning of the resonances between sub-systems – potentially reaching a full FSR. In view of the above, it is an open question whether signatures of PT-symmetry are retained in such large scale settings.

Here, we demonstrate that many features of PT-symmetry are sufficiently robust so as to survive the statistical fluctuations associated with macroscopic fiber cavities – even ones having a length in excess of 1 km. Starting from a generic linear PT-symmetric laser cavity model, we construct a conceptually analogous lumped-component model that we experimentally realize in a single-mode-fiber cavity. Coherent coupling and feedback from the interfaces in the traditional model are replaced by partially reflective fiber Bragg mirrors connecting two sub-cavities in which optical amplification and attenuation are provided by localized components in lieu of the distributed gain and loss used in previous approaches. In such an arrangement, the gain-loss balance is readily maintained and varied deterministically, but the sub-cavity phases *cannot* be held fixed due to unavoidable fluctuations in such a large system – thereby leading to resonance detuning. Nevertheless, we demonstrate experimentally and theoretically that the lasing threshold and the post-lasing output-power scaling in the PT-symmetric configuration survives the statistical detuning effects of the sub-cavity resonances – even when this detuning spans the full FSR. We present the first *quantitative* identification of lasing thresholds and broken and unbroken PT-symmetric lasing phases, which is made possible by the unambiguous separation of the power emitted by the gain and loss sub-cavities. Furthermore, we find that although detuning precludes the existence of an *exact* unbroken PT-symmetric phase, observation of the signature of symmetry breaking is nevertheless enabled through tuning the attenuation of the loss sub-cavity. The demonstrated robustness of PT-symmetry in macroscopic fiber systems could pave the way to applications of such concepts in telecommunications and fiber lasers.

A complete elimination of detuning between coupled resonators is obviously not possible due to fabrication limitations. However, micro-scale settings afford the benefit of not only restricting the detuning to small values in comparison with the cavity free spectral range but also of having the detuning invariant over time. We deliberately dispense with such near-ideal configurations in order to study non-Hermitian effects in a statistical system. In the coupled fiber arrangement considered here, the detuning between the cavity resonances is of the order of the free spectral range and also fluctuates randomly over time.

To study non-Hermitian phenomena in large scale settings, we first theoretically analyze the ideal zero-detuned cavity geometry using a mean-field saturable model to obtain two distinct steady-state nonlinear supermodes. We then find that even in the presence of a random detuning, the system behavior still mimics this zero-detuned ideal scenario. In this case, although the supermodes are not formally equivalent to those when the detuning is absent, the existence of unbroken and broken PT-symmetry phases in this statistical system can still be inferred. Moreover, we show that a gradual phase transition can also take place despite the random fluctuations between the cavity resonances. In order to observe this transition, we move away from the PT-symmetric condition of balancing the gain and loss. As the loss is constantly increased, a transition occurs from the phase of decaying lasing power to the one with an increase in the lasing power from the gain side. We attribute this aspect to the presence of an exceptional point in this non-Hermitian arrangement. Although the detuning fluctuates over time, our results indicate that, on average, the system can still operate in the vicinity of the exceptional point and thereby display the aforementioned phase transition. Our results thus extend the manifestation of non-Hermitian effects in large scale non-deterministic settings.

## Results

**Lumped-component model of a photonic PT-symmetric system.** We start by abstracting from an archetypical optical PT-symmetric configuration (Fig. 1a) an equivalent discrete 'lumped-component' system (Fig. 1b-d). The arrangement shown in Fig. 1a consists of equal lengths of homogeneous materials of refractive indices $n_g$ and $n_\ell$ in intimate contact and surrounded with symmetric external media. The imaginary part of the index corresponds to either optical loss ($n_\ell$) or gain ($n_g$), depending on its sign. If the indices satisfy $n_g^* = n_\ell$, then the structure is said to be PT-symmetric. This condition entails that the real part of the refractive index has an even distribution (it is equal in both layers), whereas the imaginary part has an odd distribution (optical gain in one layer and matching losses in the other). Index discontinuities at all three interfaces provide reflection that is particularly weak at the interface between the two layers (where it depends on only the contrast between the imaginary components of $n_g$ and $n_\ell$) – resulting in strong coupling between the two layers. Despite the simplicity of this fundamental model, it has not been experimentally realized to date – in part due to the constraints placed by the Kramers-Kronig relationships on the commensurate values of the real and imaginary components of the refractive index of any material [44]. To date, many physical realizations of PT-symmetric cavities have focused instead on other micro-systems such as coupled ring cavities or parallel waveguides.

The optical structure shown in Fig. 1b that comprises two coupled sub-cavities is conceptually equivalent to that in Fig. 1a. Fresnel reflection at the interfaces is replaced by partially reflecting mirrors: outer symmetric mirrors $M_1$ and $M_3$ having equal reflectivities $R_1 = R_3 = R$ that correspond to the interfaces with the external media, and a middle mirror $M_2$ of reflectivity $R_2$ that couples the two sub-cavities and corresponds to the interface between the gain and loss layers. Discrete optical amplifiers and attenuators provide single-pass amplification $G$ and attenuation $\mathcal{L}$ in the sub-cavities. Crucially, in such a configuration the reflections are no longer constrained by the physical limitations on the refractive indices of materials as dictated by the Kramers-Kronig relationships. Instead, coherent feedback between the sub-cavities becomes *independent* of the gain/loss contrast.

In the lumped-component model, balanced gain and loss corresponds to $G\mathcal{L} = 1$, removing attenuation altogether from the lossy sub-cavity corresponds to $\mathcal{L} = 1$ (gain-lossless configuration in Table 1), and an infinite attenuation to $\mathcal{L} = 0$ (gain sub-cavity in Table 1). From this perspective, the PT-symmetric arrangement is a particular element in a continuum of possibilities where $\mathcal{L}$ is varied and $G$ is held fixed. This family of structured laser cavities can be viewed as a result of inserting passive elements (the mirror $M_2$ and the attenuation $\mathcal{L}$) into a symmetric reference cavity consisting of an amplifying gain element $G$ between two mirrors $M_1$ and $M_3$ having an equal reflectivity $R$ (Fig. 1a,b; reference gain cavity) – corresponding to the addition of a loss layer to the gain layer. A question naturally arises whether the establishment of PT-symmetry by inserting a gain-balancing loss will inevitably raise the lasing threshold of this structured cavity with respect to that of a reference cavity where the loss is eliminated? We proceed to show that this is *not* always the case.

**Experimental realization in a macroscopic PT-symmetric fiber-based cavity.** To realize the lumped-component PT-symmetric structure shown in Fig. 1b, we have constructed a C-band single-mode-fiber-based cavity in which all the degrees of freedom are independently accessible, as illustrated in Fig. 1c. Gain is produced by a fiber-pigtailed semiconductor optical amplifier (SOA), the loss is induced by a variable optical attenuator (VOA), and optical feedback is provided by custom-made fiber Bragg gratings (FBGs) with desired reflectivity, central wavelength, and bandwidth (Methods). A single polarization is maintained by utilizing a polarization-sensitive SOA and polarization-maintaining optical components. Here we keep the reflectivities of the side mirrors fixed at $R \approx 82\%$ (left and right external FBGs $M_1$ and $M_3$), and vary $R_2$ from 7% to 99% for the intra-sub-cavity coupling FBG $M_2$.

We first measure the lasing threshold of the PT-symmetric configuration by gradually increasing the contrast between $G$ and $\mathcal{L}$ while maintaining the balanced condition $G\mathcal{L} = 1$ until lasing is initiated. The lasing thresholds $G = G_{\text{PT}}$ for different $R_2$ are listed in Table 1. Measurements of the thresholds at the two limits of the above-described continuum of arrangements while varying $\mathcal{L}$ are also listed. At $\mathcal{L} = 1$ (gain-lossless), the threshold $G_0$ is always less than $G_{\text{PT}}$; whereas for $\mathcal{L} = 0$ (gain sub-cavity), the threshold $G_{\text{open}}$ is always higher than $G_{\text{PT}}$. The PT-symmetric cavity may have higher or lower threshold $G_{\text{PT}}$ in comparison to the threshold $G_{\text{Ref}}$ for the reference cavity (after removing the passive elements $M_2$ and $\mathcal{L}$). Measurements of the thresholds in all four cavity configurations while varying $R_2$ are compared to theoretical values obtained by identifying the poles of the cavity transmission (Supplementary Sec. S1),

$$G_{\text{PT}} = \frac{1-R}{2\tilde{R}} + \sqrt{1 + \left(\frac{1-R}{2\tilde{R}}\right)^2}, \; G_0 = \frac{1+\tilde{R}}{R+\tilde{R}}, \; G_{\text{open}} = \frac{1}{\tilde{R}}, \; G_{\text{Ref}} = \frac{1}{R}, \tag{1}$$

where $\tilde{R} = \sqrt{RR_2}$. These expressions remain unaffected whether the cavities are deterministic or if we assume randomly varying phases inserted in the sub-cavities.

We would like to stress the reason behind comparing $G_{\text{PT}}$ with $G_{\text{Ref}}$ and not the other 3-mirror systems. In most of the models considered in previous works that deal with coupled disk or cavity lasers, the inter-cavity coupling exists because of the evanescent tails of the fields in the cavities. In this scenario, once one sub-cavity is removed from the system (for example the lossy cavity), the coupling also disappears. Comparing this to our linear cavity arrangement, the coupling is provided by the central mirror

$R_2$. Hence, in order to remove the loss altogether, a fair comparison can only be made if the coupling mirror is also removed, which is the reason for comparing $G_{PT}$ with $G_{Ref}$.

The differences between the thresholds of the distinct cavity configurations listed in Table 1 are most prominent at low $R_2$, whereupon the two sub-cavities are strongly coupled. As $R_2$ increases, the differences between $G_{PT}$, $G_0$, and $G_{open}$ are reduced monotonically and ultimately vanish as the amplifying sub-cavity is effectively isolated from its attenuating counterpart. In comparing $G_{PT}$ to $G_{Ref}$, however, we find two regimes while varying $R_2$. At low $R_2$ (strong coupling), we have $G_{PT} > G_{Ref}$, whereas increasing $R_2$ can result in $G_{PT} < G_{Ref}$, therefore indicating that introducing gain-balancing loss into the reference cavity – counter-intuitively – may help reduce the lasing threshold. The advantage of a PT-symmetric cavity over the reference gain cavity is brought out in Fig. 2 where we plot the threshold-reduction factor $\eta = G_{Ref}/G_{PT}$. This ratio is unity when $R_2 = R/(1+R)^2$, which is identified by the black curve in Fig. 2 that divides the parameter space into two regions: $\eta > 1$ where PT-symmetry helps lower the lasing threshold, and $\eta < 1$ where it does not. The lasing threshold is reduced despite introducing gain-balancing loss in the cavity whenever $R_2$ and $R$ are judiciously selected.

**Theoretical model for the coupled-cavity laser system.** The expressions for lasing thresholds in Eq. (1) are obtained via the transfer matrix method that posits a linear model for all the optical components (Supplementary Sec. S1). As such, this approach is not suitable for describing the lasing dynamics whereupon the fields may experience exponential growth. In this nonlinear regime, we employ a mean-field temporal coupled-mode approach [45] (Supplementary Sec. S2) in which the averaged field amplitudes in the gain sub-cavity $a$ and the loss sub-cavity $b$ are coupled through

$$\frac{da}{dt} = -\gamma_1 a + i\frac{\Delta}{2}a + \frac{g}{1+|a|^2}a + i\kappa b, \qquad (2)$$

$$\frac{db}{dt} = -\gamma_2 b - i\frac{\Delta}{2}b + i\kappa a. \qquad (3)$$

Here we have introduced an effective temporal coupling coefficient $\kappa$ between the sub-cavities (to be defined below); $\gamma_1$ and $\gamma_2$ are temporal linear losses in the amplifying and attenuating sub-cavities, respectively (Supplementary Eq. S2.1 and Eq. S2.2), which incorporate leakage from the side mirrors and the loss imposed by the VOA; and $g$ is the small-signal gain. $\Delta$ is the frequency difference between the resonances of the sub-cavities (Fig. 3a-b); henceforth referred to as the 'detuning'. These parameters are all related to the mirror reflectivities and fiber lengths (Supplementary Sec. S2). We introduce gain saturation in Eq. 2 to capture the power dynamics after the onset of lasing [42]. A useful feature of this model is that it can apply to a wide range of non-Hermitian photonic systems beyond ours.

We are still missing a model for the temporal coupling coefficient $\kappa$ between the two sub-cavities. To this end, we have adapted the Lamb coupled-cavity model [46,47]. Lamb modeled the coupling between two coupled cavities through a permittivity 'bump' of infinitesimal thickness, which we relate here to the reflectivity $R_2$. Additionally, leakage from the finite-reflectivity side mirrors effectively produces a shift in the FSR, leading to a dependence of $\kappa$ on $R$. These considerations lead to an expression for $\kappa$,

$$\kappa = \frac{v_g}{2n_o^2 d}(1-R)\sqrt{\frac{1-R_2}{R_2}}, \qquad (4)$$

where $d$ is the fiber-cavity length, $v_g$ is the group velocity, and $n_o$ is the refractive index (see Supplementary Sec. S3 for details).

In light of the macroscopic nature of the fiber-based cavity, we assume that the detuning $\Delta$ is a random variable. Indeed, given the long cavity length, and thus the extremely small free spectral range (FSR), slight perturbations in the experimental conditions may cause $\Delta$ to potentially vary across the whole FSR. The solutions are obtained numerically by carrying out an ensemble average over a distribution for $\Delta$, either a Gaussian distribution $P(\Delta) \propto \exp\{-\Delta^2/(2\sigma^2)\}$ (Fig. 3c) or a uniform distribution (Fig. 3d) as candidate models. Analysis under these considerations leads to the intriguing conclusion that features associated with the presence of an exceptional point (a non-Hermitian degeneracy) [48-54] can – in principle – still be detected.

**Linear coupled-cavity model for predicting the lasing threshold and symmetry-breaking condition.** The lasing modes can be obtained from Eqs. 2-3 in the steady-state. This model is valid both before and after lasing occurs, and to ensure the consistency of our analysis we compare computed lasing thresholds of the PT-symmetric arrangement to those obtained from the linear transfer matrix method (Eq. 1). To achieve this, we linearize Eq. 2 by ignoring gain saturation and set the detuning to $\Delta = 0$, and then determine the lasing thresholds by assuming a harmonic ansatz for Eqs. 2-3 of the form $\begin{pmatrix} a(t) \\ b(t) \end{pmatrix} = \begin{pmatrix} a_0 \\ b_0 \end{pmatrix} e^{i\lambda t}$, where $\begin{pmatrix} a_0 \\ b_0 \end{pmatrix}$ is a constant vector. The general solution for the eigenvalues has the form

$$\lambda_{1,2} = -\frac{i}{2}(g - \gamma_1 - \gamma_2) \pm \kappa\sqrt{1 - \left(\frac{g-\gamma_1+\gamma_2}{2\kappa}\right)^2}. \tag{5}$$

Within the linear, zero-detuning, PT-symmetric configuration ($g + \gamma_1 = \gamma_2$), the two eigenvalues are $\lambda_{1,2} = i\gamma_1 \pm \sqrt{\kappa^2 - g^2}$. The lasing threshold is identified by determining the onset for a negative imaginary component of the eigenvalues, $g_{\text{th}} = \sqrt{\gamma_1^2 + \kappa^2}$. Computing the lasing threshold based on this model reveals excellent agreement with the predictions of the transfer matrix method for the PT-symmetric structure (Supplementary Fig. S3).

The behavior of the eigenvalues while varying $g$ displays a bifurcation, as illustrated in Fig. 3e-f (dashed curves). When $g < \kappa$, the eigenvalues have the same imaginary part $i\gamma_1$ but distinct real parts $\pm\sqrt{\kappa^2 - g^2}$. As $g \to \kappa$, the real parts coalesce at zero (Fig. 3e) whereas the imaginary components diverge along forked trajectories (Fig. 3f). We denote the range $g < \kappa$ as the 'unbroken' PT-symmetry regime (U), and the range $g > \kappa$ the 'broken' PT-symmetry regime (B), separated by the exceptional point at $g = \kappa$. The behavior of the field is quite distinct in these two regimes. The unbroken-PT regime features equal field amplitudes in the two sub-cavities $\begin{pmatrix} a_0 \\ b_0 \end{pmatrix} = \begin{pmatrix} 1 \\ \pm e^{\pm i\theta} \end{pmatrix}$, where $\sin\theta = g/\kappa$. The power emitted from the gain and loss sub-cavity ports are thus expected to be equal. In the broken-PT regime, the modal field is more concentrated in the gain or loss sub-cavity having unequal amplitudes $\begin{pmatrix} a_0 \\ b_0 \end{pmatrix} = \begin{pmatrix} 1 \\ ie^{\pm\theta} \end{pmatrix}$, where $\cosh\theta = g/\kappa$, leading to unequal power emission from the two ports. Four sharply delineated domains of operation can be identified while varying the loss and gain *independently*: lasing in B, lasing in U, non-lasing in B, and non-lasing in U, as depicted in Fig. 3g.

We now consider the impact of detuning $\Delta$ on the system while retaining the linear PT-symmetric condition ($G\mathcal{L} = 1$). As $\Delta$ increases, the bifurcation in the real and imaginary parts of the eigenvalues is 'smeared out' in a complementary fashion. Prior to the EP, the real part closely resembles the zero-detuning results, but deviates considerably after the EP. The opposite is observed in the imaginary part: it closely follows the zero-detuning results after the EP and diverges beforehand. It can be shown on theoretical grounds that the presence of detuning precludes the observation of a pure unbroken-PT mode (Supplementary Sec. S5). We can nevertheless define a pseudo-unbroken symmetry regime, whereupon the $|a_0| \approx |b_0|$ and the amplitudes are affected in a similar manner upon changing the gain and loss [43]. Note that in a *strict* PT-symmetric configuration (the dashed zero-detuning curves in Fig. 3e-f), lasing will only occur in the broken-symmetry regime, which has been the case in previous experiments [32,33]. Nevertheless, the calculations in Fig. 3e-f show that the smearing of the bifurcation resulting from detuning can produce lasing in the unbroken-PT regime. Furthermore, this restriction can be relaxed by relying on unbalanced gain and loss ($G\mathcal{L} \neq 1$; Fig. 3g) [42].

**Nonlinear steady-state coupled-cavity model for PT-lasing dynamics.** Various models have recently been put forth to study the interplay of nonlinearity and PT-symmetry [55-58]. In the structure under consideration, the lasing field dynamics, such as power-scaling with gain, can be analyzed using the nonlinear model in Eqs. (2-3). An important property of lasing structures in the steady-state is that the saturated gain always clamps to the net amount of attenuation present in the system [59]. A critical consequence of this general physical restriction is that the gain/loss contrast no longer determines the transition between different symmetry phases, only the *loss* does. We confirm this prediction by again employing a harmonic ansatz (with $\Delta = 0$) but without imposing a balance between gain and loss. Instead, we regard them as independent variables. By allowing for only *real* eigenvalue solutions (as a result of gain clamping), we obtain analytical expressions for two distinct phases of field oscillation, which we map to the unbroken (U) and broken (B) PT-symmetry regimes described above:

$$\gamma_2 \leq \kappa: \begin{pmatrix} a \\ b \end{pmatrix}_U = \sqrt{\frac{g}{\gamma_1+\gamma_2} - 1} \begin{pmatrix} 1 \\ \pm e^{\pm i\theta} \end{pmatrix} e^{\pm i(\kappa \cos\theta)t}, \qquad (6)$$

$$\gamma_2 > \kappa: \begin{pmatrix} a \\ b \end{pmatrix}_B = \sqrt{\frac{g}{\gamma_1+\kappa^2/\gamma_2} - 1} \begin{pmatrix} 1 \\ i\kappa/\gamma_2 \end{pmatrix}, \qquad (7)$$

where the parameter $\theta$ in Eq. 6 is obtained from $\sin\theta = \gamma_2/\kappa$. Two new features emerge here. In contrast to the linear model in which the gain/loss contrast determines the boundary between the broken and unbroken regimes, this boundary in the nonlinear regime is dictated by the loss $\gamma_2$ in the lossy sub-cavity alone. The unbroken PT-phase (Eq. 6) is characterized by equal intensities $|a|^2 = |b|^2$ in the sub-cavities and the two nonlinear supermodes are split in frequency by $2\kappa \cos\theta$. On the other hand, the broken PT-phase entails an unequal distribution of intensities with $|a|^2 > |b|^2$. Another important feature is that the supermodes now exhibit fixed amplitudes because of nonlinearity, dictated by the cavity gain and loss values, in contradistinction to the linear regime. An intuitive explanation of loss induced enhancement in lasing power is based on the fact that as the loss increases, the field profile in the system becomes more asymmetric. In other words, the mode gets more localized towards the gain side, thereby leading to a rise in the lasing power $I_{\text{Gain}}$. Equation (7) showing a broken (B) nonlinear supermode, quantitatively captures this behavior since the ratio between the steady-state fields in the gain and loss cavities, i.e. $\gamma_2/\kappa$, increases as the loss $\gamma_2$ increases.

In our experiment conducted on a macroscopic fiber system extending for many meters, a pertinent question is whether the observation of such prominent broken and unbroken phases is still possible in the presence of the unavoidable resonance detunings. As a first demonstration of the validity of the nonlinear analysis described above, we measure the power-scaling characteristics of the PT-symmetric laser while holding $R_2 = 6.8\%$ fixed and increasing the gain-loss contrast while maintaining the balance $G\mathcal{L} = 1$. A unique feature of our experimental arrangement is that the output power from the loss and gain sub-cavity ports ($I_{\text{Gain}} = |a|^2$ and $I_{\text{Loss}} = |b|^2$) can be recorded separately and quantitatively (Fig. 4). It is thus possible to determine unambiguously whether lasing is initiated in the broken or unbroken symmetry phases. We carried out these measurements in two cavity configurations that we denote 'short' and 'long'. In the short cavity, the total length is $d \approx 6$ m, which is associated with a FSR of $\lambda_{\text{FSR}} = \lambda^2/2nd = 133$ fm. The cavity has a quality factor of $Q = 5.2 \times 10^7$ and a finesse of $\mathcal{F} = 14$. In the long cavity, we inserted an extra 1-km-long fiber spool in the loss sub-cavity (Fig. 1d), which exacerbates the detuning between the two sub-cavities. The total length is $d \approx 1$ km, the FSR is $\lambda_{\text{FSR}} = 0.8$ fm, $Q = 8.7 \times 10^9$, and $\mathcal{F} = 14$. The data reveals clearly that lasing occurs in the broken regime $I_{\text{Gain}} \neq I_{\text{Loss}}$ in both cavity configurations. Note however that $I_{\text{Gain}} \approx I_{\text{Loss}}$ at low gain/loss contrasts, which indicates that an unbroken phase is approached, as can be expected from Fig. 3g.

To compare the data on power-scaling with predicted values based on the nonlinear model, we must include the impact of phase detuning in the system of Eqs. 2-3. We compute an ensemble average over a Gaussian distribution for $\Delta$ over one free spectral range; Fig. 3c. The standard deviation $\sigma$ plays an important role in determining the lasing characteristics. We fitted the results of the coupled model for different values of $\sigma$ and obtained a good match for $\sigma = \omega_{FSR}/10$. This quantifies the amount of average resonance detuning between the two coupled fiber sub-cavities. Utilizing a uniform probability distribution (Fig. 3d) predicts a substantially larger contrast between $I_{\text{Gain}}$ and $I_{\text{Loss}}$ than that observed experimentally.

The trends in Fig. 4 clearly show that the disparity between $I_{\text{Gain}}$ and $I_{\text{Loss}}$ continues to grow with $g$, thus confirming that the mode in the gain sub-cavity further localizes as the gain-loss contrast in the PT-system is enhanced [5,15,32-35]. This is a well-known feature of the broken-PT phase. Since the steady-state always remains in this phase for the balanced values of $\gamma_2 = g + \gamma_1$ maintained here, we deduce from the results in Eq. 2 that the range over which the loss is varied is actually higher than the coupling strength between the fiber cavities for $R_2 = 7\%$.

**Observing statistical PT-symmetry breaking.** Finally, we demonstrate that our macroscopic fiber-based laser-cavity system – despite the extreme random detuning between the sub-cavities – still displays the signature of an exceptional point. It is clear from Fig. 3g that transitioning between the unbroken- and broken-symmetry phases associated with a lasing system in the steady state can take place by varying the loss $\gamma_2$ alone at fixed gain $g$. Crucially, a quantitative observation of this transition necessitates independent tuning of the gain and loss and unambiguous measurements of the power at the two output ports. Both of these desiderata are satisfied in our experimental arrangement. We vary $\gamma_2$ via the VOA after holding $g$ as provided by the SOA fixed at a value well above the lasing threshold of the gain sub-cavity, such that lasing occurs regardless of $\gamma_2$. We have carried out this experiment for four values of single-pass amplification ($G = 15, 20, 25, 30$ dB), and for each value we sweep the VOA single-pass attenuation from 0 to 25 dB while recording the lasing power at the two output ports. Increasing the loss results in a monotonic drop in power from the loss port as might be expected (Fig. 5c-d). However, the result for the gain port is counterintuitive: the power initially drops with increasing loss, but then increases with further loss is added (Fig. 5a-b). This increase in lasing power with additional loss is particularly visible when the gain is held at 30 dB. At a gain of 15 dB, this effect has vanished and a transition is no longer detectable.

Loss-induced enhancement of lasing power has been observed in microcavities and is attributed to the notion of an exceptional point. While the same effect is observed in this macroscopic cavity, it is worth mentioning that strictly pure broken and unbroken phases do not exist in this cavity because the propagation phases are not deterministic. Yet, even in this statistical environment, we have confirmed for the first time that a PT-phase transition is still observable in such large-scale active cavities.

We note that a perfect eigenstate coalescence disappears in our system as a consequence of the statistical fluctuations. But this does not have a detrimental effect on the expected response of the system: the behavior of eigenvalues just gets smoothed out when the loss $\gamma_2$ is varied. In other words, instead of coalescing, the eigenvalues still approach each other around the original location of the EP (in the zero-detuned case) and also tend to bifurcate afterwards. This is the reason why a loss induced enhancement of lasing power is still observable in our randomly-varying system (see Supplementary S6 for details).

## Discussion

Early studies of PT-symmetry focused solely on deterministic models, owing to the micro-scale nature of the coupled resonators and waveguides investigated. In such settings, the impact of a fixed deterministic detuning can be easily understood. In contrast to this scenario, the random phase fluctuations in macroscopic-scale photonic systems lead to drastic detuning between coupled cavities that may span in principle the full free spectral range (FSR). Carrying out an ensemble average over a range of outcomes becomes necessary, and it is not clear *a priori* that the essential features of PT-symmetry will survive.

In this work, we have presented the first realization of a statistical PT-symmetric lasing structure using coupled fiber cavities extending over a kilometer in length. An important outcome of our experiment has been the persistence of many essential features of the deterministic formulation of PT-symmetry. For example, the lasing threshold was found to be robust against random phase fluctuations. Indeed, the lasing threshold in a PT-symmetric cavity containing a gain-balancing loss can be lower than that of the same cavity after removing the loss and the coupling mirror, thus potentially providing significant benefits compared to a gain-only cavity. In all cases, the experimental results are in agreement with theoretical predictions after ensemble averaging over a Gaussian distribution of detuning values.

With regards to lasing power emerging from the gain and loss ports of our structure, we have observed a transition between the two well-known phases of unbroken and broken symmetry in this statistical system occurring around an exceptional point. The presence of random phase fluctuations, however, prevents a complete coalescence of eigenvalues and thus precludes the observation of a pure unbroken phase where the lasing powers from the gain and loss ports are equal. Crucially, in this nonlinear system the transition behavior between the unbroken- and broken-symmetry phases is dictated only by the loss in the lossy sub-cavity. Despite the statistical nature of the experimental arrangement, the optical power decays in unison from both the loss and gain cavities with increasing cavity loss until the exceptional point is reached, after which the power counterintuitively begins to rise at the gain port with further increase in the incorporated loss. Such loss-induced transparency and lasing effects have so far been observable only in micro-scale devices. Our results thus indicate that the notion of PT-symmetry – and non-Hermitian optics in general – may have impact on large scale non-deterministic platforms such as fiber networks.

## Methods

**Experimental arrangement.** In the fiber-based cavity, the mirrors are custom-made FBGs on single-mode fibers (SMF28) in the C-band (O-Eland Inc., central wavelength $\approx$ 1552.5 nm, bandwidth $\approx$ 5 nm). The gain of SOA is fine-tuned (resolution< 0.05 dB). Similar fine-tuning for optical attenuation is achieved by cascading the VOA (Thorlabs VOA50PM-APC) with a secondary SOA (Thorlabs BOA1004P), which enables high-resolution adjustment of the net loss in the lossy sub-cavity. The gain spectra of the SOAs were calibrated over the bandwidth of operation (5 nm) using a tunable laser (Agilent 8164A Mainframe with 81680A Tunable Laser Source Module). The VOA employed in our setup has a flat spectrum since it induces loss by physically blocking the optical beam. All fibers pigtails are APC-type to minimize unwanted reflections at the fiber connections. All the SOAs operate in the TE-polarization mode, and we ensure that the polarization of the fiber mode is TE when it reaches the SOAs. In all the experimental configurations (except the long cavity), the fibers are polarization-maintaining and the beam polarization remains TE when circulating throughout the cavity. For the long cavity, we use polarization controllers to guarantee that TE-polarization reaches the SOAs. The output power from each port is recorded by an optical spectrum analyzer (OSA, Advantest Q8381A) with a spectral resolution of 0.1 nm.

The simulation results provided in Figs. 4 and 5 for the nonlinear system of Eqs. (2-3) were carried out by first finding steady-state values of $|a|^2$ and $|b|^2$ for a specific value of the detuning $\Delta$. We then compute an ensemble average for these steady-state intensities over a full FSR assuming either a Gaussian or a uniform probability distribution for $\Delta$. This is explained in Supplementary section S7, and the equation involved is stated here for convenience of the reviewer:

$$\langle I_{a,b}^{(ss)} \rangle = \int_{-\omega_{\text{FSR}}/2}^{\omega_{\text{FSR}}/2} I_{a,b}^{(ss)}(\Delta) P(\Delta) d\Delta \qquad (8)$$

Here $P(\Delta)$ is the probability distribution followed by the detuning $\Delta$ and $I_{a,b}^{(ss)}$ is the steady state intensity found for a specific value of $\Delta$. To obtain the final curves given in Figs. 4 and 5, this procedure is followed for each parameter value over the whole range of the loss $\gamma_2$ and gain $g$.

## References

[1] C. M. Bender and S. Boettcher, "Real spectra in non-Hermitian Hamiltonians having PT-symmetry," Phys. Rev. Lett. **80**, 5243–5246 (1998).

[2] C. M. Bender, "Making sense of non-Hermitian Hamiltonians," Rep. Prog. Phys. **70**, 947–1018 (2007).

[3] R. El-Ganainy, K. G. Makris, D. N. Christodoulides, and Z. H. Musslimani, "Theory of coupled optical PT-symmetric structures," Opt. Lett. **32**, 2632–2634 (2007).

[4] K. G. Makris, R. El-Ganainy, D. N. Christodoulides, and Z. H. Musslimani, "Beam dynamics in PT symmetric optical lattices," Phys. Rev. Lett. **100**, 103904 (2008).

[5] C. E. Rüter, K. G. Makris, R. El-Ganainy, D. N. Christodoulides, M. Segev, and D. Kip, "Observation of parity–time symmetry in optics," Nat. Phys. **6**, 192–195 (2010).

[6] O. Bendix, R. Fleischmann, T. Kottos, and B. Shapiro, "Exponentially fragile PT symmetry in lattices with localized eigenmodes," Phys. Rev. Lett. **103**, 030402 (2009).

[7] S. Longhi, "Bloch oscillations in complex crystals with PT symmetry," Phys. Rev. Lett. **103**, 123601 (2009).

[8] C. T. West, T. Kottos, and T. Prosen, "PT-symmetric wave chaos," Phys. Rev. Lett. **104**, 054102 (2010).

[9] E.-M. Graefe and H. F. Jones, "PT-symmetric sinusoidal optical lattices at the symmetry-breaking threshold," Phys. Rev. A **84**, 013818 (2011).

[10] R. Fleury, D. Sounas, and A. Alù, "An invisible acoustic sensor based on parity-time symmetry," Nat. Commun. **6**, 5905 (2014).

[11] X. Zhu, H. Ramezani, C. Shi, J. Zhu, and X. Zhang, "PT-symmetric acoustics," Phys. Rev. X **4**, 031042 (2014).

[12] Hui Jing, S. K. Özdemir, Xin-You Lü, Jing Zhang, Lan Yang, and Franco Nori, "PT-symmetric phonon laser," Phys. Rev. Lett. **113**, 053604 (2014).

[13] J. Schindler, A. Li, M. C. Zheng, F. M. Ellis, and T. Kottos, "Experimental study of active LRC circuits with PT symmetries," Phys. Rev. A **84**, 040101(R) (2011).

[14] J. Schindler, Z. Lin, J. M. Lee, H. Ramezani, F. M. Ellis, and T. Kottos "PT-symmetric electronics," J. Phys. A **45**, 444029 (2012).

[15] A. Guo, G. J. Salamo, D. Duchesne, R. Morandotti, M. Volatier-Ravat, V. Aimez, G. A. Siviloglou, and D. N. Christodoulides, "Observation of PT-symmetry breaking in complex optical potentials," Phys. Rev. Lett. **103**, 093902 (2009).

[16] M. Brandstetter, M. Liertzer, C. Deutsch, P. Klang, J. Schöberl, H. E. Türeci, G. Strasser, K. Unterrainer, and S. Rotter, "Reversing the pump dependence of a laser at an exceptional point," Nat. Commun. **5**, 4034 (2014).

[17] B. Peng, Ş. K. Özdemir, S. Rotter, H. Yilmaz, M. Liertzer, F. Monifi, C. M. Bender, F. Nori, and L. Yang, "Loss-induced suppression and revival of lasing," Science **346**, 328–332 (2014).

[18] A. Regensburger, C. Bersch, M. A. Miri, G. Onishchukov, D. N. Christodoulides, and U. Peschel, "Parity–time synthetic photonic lattices," Nature **488**, 167–171 (2012).

[19] M. Wimmer, A. Regensburger, M.-A. Miri, C. Bersch, D. N. Christodoulides, and U. Peschel, "Observation of optical solitons in PT-symmetric lattices," Nat. Commun. **6**, 7782 (2015).

[20] S. Longhi, "PT-symmetric laser absorber," Phys. Rev. A **82**, 031801 (2010).

[21] Z. J. Wong, Y.-L. Xu, J. Kim, K. O'Brien, Y. Wang, L. Feng, and X. Zhang, "Lasing and anti-lasing in a single cavity," Nat. Photon. **10**, 796–801 (2016).

[22] G. Castaldi, S. Savoia, V. Galdi, A. Alù, and N. Engheta, "PT metamaterials via complex-coordinate transformation optics," Phys. Rev. Lett. **110**, 173901 (2013).

[23] N. Lazarides and G. P. Tsironis, "Gain-driven discrete breathers in PT-symmetric nonlinear metamaterials," Phys. Rev. Lett. **110**, 053901 (2013).

[24] H. Alaeian and J. A. Dionne, "Parity-time-symmetric plasmonic metamaterials," Phys. Rev. A **89**, 033829 (2014).

[25] M. Lawrence, N. Xu, X. Zhang, L. Cong, J. Han, W. Zhang, and S. Zhang, "Manifestation of PT symmetry breaking in polarization space with terahertz metasurfaces," Phys. Rev. Lett. **113**, 093901 (2014).

[26] H. Zhao, W. S. Fegadolli, J. Yu, Z. Zhang, L. Ge, A. Scherer, and L. Feng, "Metawaveguide for Asymmetric Interferometric Light-Light Switching," Phys. Rev. Lett. **117**, 193901 (2016).

[27] S. Xiao, J. Gear, S. Rotter, and J. Li, "Effective PT-symmetric metasurfaces for subwavelength amplified sensing," New J. Phys. **18**, 085004 (2016).

[28] Z. Lin, H. Ramezani, T. Eichelkraut, T. Kottos, H. Cao, and D. N. Christodoulides "Unidirectional invisibility induced by PT-symmetric periodic structures," Phys. Rev. Lett. **106**, 213901 (2011).

[29] L. Feng, Y.-L. Xu, W. S. Fegadolli, M.-H. Lu, J. E. B. Oliveira, V. R. Almeida, Y.-F. Chen, and A. Scherer, "Experimental demonstration of a unidirectional reflectionless parity-time metamaterial at optical frequencies," Nat. Mater. **12**, 108–113 (2013).

[30] H. Ramezani, H.-K. Li, Y. Wang, and X. Zhang, "Unidirectional spectral singularities," Phys. Rev. Lett. **113**, 263905 (2014).

[31] H. Ramezani, Y. Wang, X. Zhang, "Unidirectional Perfect Absorber," IEEE J. Sel. Top. Quantum Electron. **22,** 115–120 (2016).

[32] H. Hodaei, M. A. Miri, M. Heinrich, D. N. Christodoulides, and M. Khajavikhan, "Parity-time–symmetric microring lasers," Science **346**, 975–978 (2014).

[33] L. Feng, Z. J. Wong, R. M. Ma, Y. Wang, and X. Zhang, "Single-mode laser by parity-time symmetry breaking," Science **346**, 972–975 (2014).

[34] L. Chang, X. Jiang, S. Hua, C. Yang, J. Wen, L. Jiang, G. Li, G. Wang, and M. Xiao, "Parity–time symmetry and variable optical isolation in active-passive-coupled microresonators," Nat. Photon. **8**, 524–529 (2014).

[35] B. Peng, S. K. Özdemir, F. Lei, F. Monifi, M. Gianfreda, G. L. Long, S. Fan, F. Nori, C. M. Bender, and L. Yang, "Parity-time-symmetric whispering-gallery microcavities," Nat. Phys. **10**, 394–398 (2014).

[36] M. C. Rechtsman, J. M. Zeuner, Y. Plotnik, Y. Lumer, D. Podolsky, F. Dreisow, S. Nolte, M. Segev, and A. Szameit, "Photonic Floquet topological insulators," Nature **496**, 196–200 (2013).

[37] L. Lu, J. D. Joannopoulos, and M. Soljačić, "Topological photonics," Nat. Photon **8**, 821–829 (2014).

[38] S. Weimann, M. Kremer, Y. Plotnik, Y. Lumer, S. Nolte, K. G. Makris, M. Segev, M. C. Rechtsman, and A. Szameit, "Topologically protected bound states in photonic parity-time-symmetric crystals," Nat. Mater., advance online publication (2016).

[39] Jan Wiersig, "Enhancing the sensitivity of frequency and energy splitting detection by using exceptional points: Application to microcavity sensors for single-particle detection," Phys. Rev. Lett. **112**, 203901 (2014).

[40] J. Li, R. Yu, C. Ding, and Y. Wu, "PT-symmetry-induced evolution of sharp asymmetric line shapes and high-sensitivity refractive index sensors in a three-cavity array," Phys. Rev. A **93**, 023814 (2016).

[41] Z.-P. Liu, J. Zhang, Ş. K. Özdemir, B. Peng, H. Jing, X.-Y. Lü, C.-W. Li, L. Yang, F. Nori, and Y.-X. Liu. "Metrology with PT-symmetric cavities: Enhanced sensitivity near the PT-phase transition," Phys. Rev. Lett. **117**, 110802 (2016).

[42] A. Cerjan and S. Fan, "Eigenvalue dynamics in the presence of nonuniform gain and loss," Phys. Rev. A **94**, 033857 (2016).

[43] A. U. Hassan, H. Hodaei, M.-A. Miri, M. Khajavikhan, and D. N. Christodoulides, "Nonlinear reversal of the PT-symmetric phase transition in a system of coupled semiconductor microring resonators," Phys. Rev. A **92**, 063807 (2015).

[44] B. E. A. Saleh and M. C. Teich, *Introduction to Photonics* (Wiley, 2007).

[45] B. E. Little, S. T. Chu, H. A. Haus, J. Foresi, and J.-P. Laine, "Microring resonator channel dropping filters," J. Lightwave Technol. **15**, 998–1005 (1997).

[46] M. B. Spencer and W. E. Lamb, "Laser with a transmitting window," Phys. Rev. A **5**, 884–892 (1972).

[47] M. B. Spencer and W. E. Lamb, "Theory of two coupled lasers," Phys. Rev. A **5**, 893–898 (1972).

[48] T. Kato, *Perturbation Theory for Linear Operators* (Springer, Berlin, 1966).

[49] Heiss, W. D. "Phases of wave functions and level repulsion," Euro. Phys. J. D **7,** 1–4 (1999).

[50] M. V. Berry, "Physics of non-Hermitian degeneracies," Czech. J. Phys. **54**, 1039–1047 (2004).

[51] S. Klaiman, U. Günther, and N. Moiseyev, "Visualization of branch points in PT-symmetric waveguides," Phys. Rev. Lett. **101**, 080402 (2008).

[52] Y. D. Chong, L. Ge, and A. D. Stone, "PT-Symmetry Breaking and Laser-Absorber Modes in Optical Scattering Systems," Phys. Rev. Lett. **106**, 093902 (2011).

[53] M. Liertzer, Li Ge, A. Cerjan, A. D. Stone, H. E. Türeci, and S. Rotter, "Pump-Induced Exceptional Points in Lasers," Phys. Rev. Lett. **108**, 173901 (2012).

[54] B. Zhen, C. W. Hsu, Y. Igarashi, L. Lu, I. Kaminer, A. Pick, S.-L. Chua, J. D. Joannopoulos, and M. Soljacic, "Spawning rings of exceptional points out of Dirac cones," Nature **525**, 354–358 (2015).

[55] A. U. Hassan, H. Hodaei, M.-A. Miri, M. Khajavikhan, and D. N. Christodoulides, "Integrable nonlinear parity-time symmetric optical oscillator," Phys. Rev. E **93**, 042219 (2016).

[56] I. V. Barashenkov and M. Gianfreda, "An exactly solvable PT-symmetric dimer from a Hamiltonian system of nonlinear oscillators with gain and loss," J. Phys. A **47**, 282001 (2014).

[57] Y. Lumer, Y. Plotnik, M. C. Rechtsman, and M. Segev, "Nonlinearly induced PT-transition in photonic systems," Phys. Rev. Lett. **111**, 263901 (2013).

[58] V. V. Konotop, J. Yang, and D. A. Zezyulin, "Nonlinear waves in PT-symmetric systems," Rev. Mod. Phys. **88**, 035002 (2016).

[59] G. Agrawal and N. Dutta, *Long-wavelength Semiconductor Lasers* (Van Nostrand Reinhold, New York, 1986).

**Acknowledgments**. This work was funded by the U.S. Air Force Office of Scientific Research (AFOSR) under MURI contract FA9550-14-1-0037, and in part by the Penn State University MRSEC through the MRSEC Program of the National Science Foundation under award number DMR-1420620.

**Tables**

**Table 1 | Comparison of the lasing threshold of a PT-symmetric cavity with those for other configurations after removing the loss element.** Four cavity models are considered here: **a**, a PT-symmetric cavity comprising sub-cavities with balanced gain and loss $G\mathcal{L} = 1$; **b**, the same cavity configuration in (**a**) after removing the loss component from its sub-cavity; **c**, a reference cavity with symmetric mirror reflectivities $R$ containing only a gain element with amplification $G$. In all cases, we examine the measured and expected lasing threshold as $R_2$ is varied. All the lasing-threshold values in the table are in dB, the value $R = 82\%$ is held fixed, and the average error in the measured thresholds is $\approx 0.2$ dB. The values of the threshold for the reference cavity (**c**) are independent of $R_2$.

| Cavity configuration | | | $R_2$ 6.8% | 18.3% | 48.5% | 99.4% |
|---|---|---|---|---|---|---|
| (a) PT-Symmetric | [R — Gain — $R_2$ — Loss — R] | $G_{PT}$ Theory | 1.71 | 1.06 | 0.66 | 0.46 |
| | | Exp. | 2.1 | 1.0 | 0.6 | 0.35 |
| (b) Gain - Lossless | [R — Gain — $R_2$ — R] | $G_0$ Theory | 0.73 | 0.64 | 0.54 | 0.46 |
| | | Exp. | 0.65 | 0.65 | 0.45 | 0.35 |
| (c) Reference Cavity | [R — Gain — R] | $G_{Ref}$ Theory | 0.86 | 0.86 | 0.86 | 0.86 |
| | | Exp. | 0.95 | 0.95 | 0.95 | 0.95 |

**Figures and figure captions**

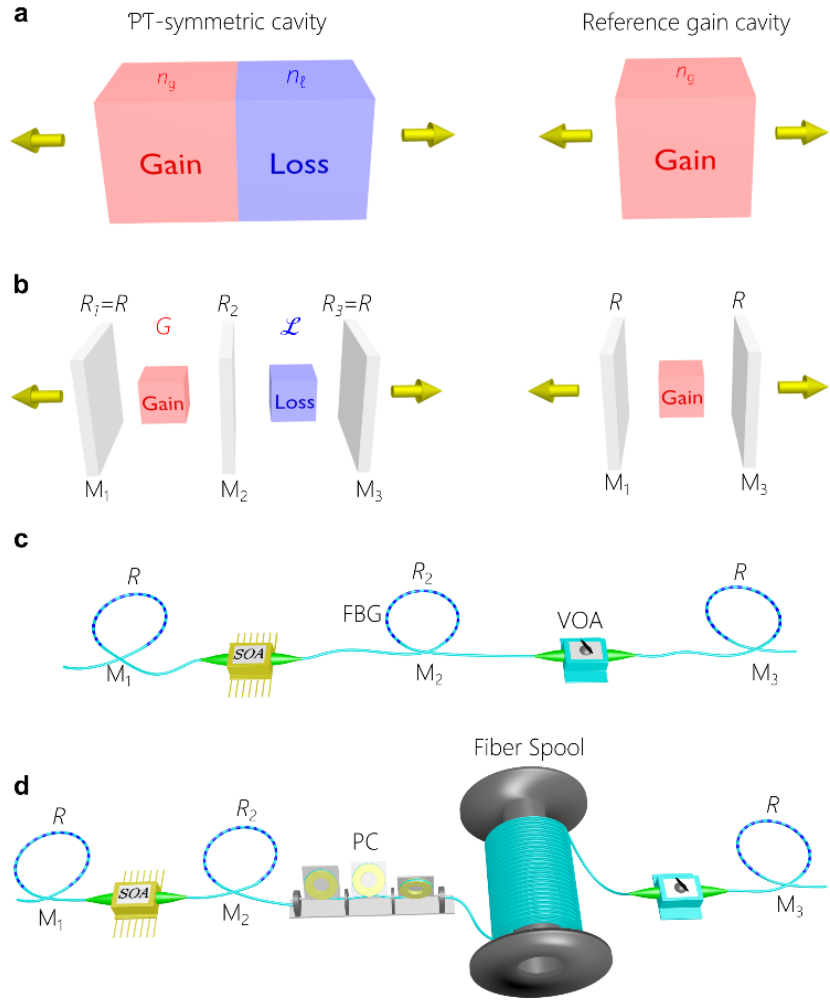

**Figure 1 | A lumped-component model of a PT-symmetric laser cavity. a**, A PT-symmetric structure formed of two homogeneous layers of refractive indices $n_g$ and $n_\ell$, (corresponding to optical gain and loss, respectively) and equal thicknesses. PT-symmetry requires $n_g = n_\ell^*$. As a reference, the gain layer alone (corresponding to the 'reference cavity' in Table 1) after removing the loss layer is shown on the right. **b**, A discrete model composed of lumped components to replace the continuum model in (**a**): the interfaces are replaced with localized mirrors, and the distributed gain and loss are replaced with an amplifier (amplification factor $G$) and an attenuator (attenuation factor $\mathcal{L}$), respectively. PT-symmetry requires that $R_1 = R_3 = R$ and $G\mathcal{L} = 1$. The cavity corresponding to the gain layer alone is formed of the side mirrors containing the amplifier. **c**, Schematic of an experimental realization of the PT-symmetric structured cavity shown on the left in (**b**) using single-mode optical fibers. Specially designed fiber Bragg grating (FBGs) are used as partially reflecting mirrors with reflectivies $R$, $R_2$, and $R$ from left to right. Gain is provided by a semiconductor optical amplifier (SOA) and attenuation by a variable optical attenuator (VOA). **d**, Optical setup in (**c**) after inserting an additional 1-km-long fiber spool. A polarization controller (PC) is added to maintain the state of polarization throughout the cavity.

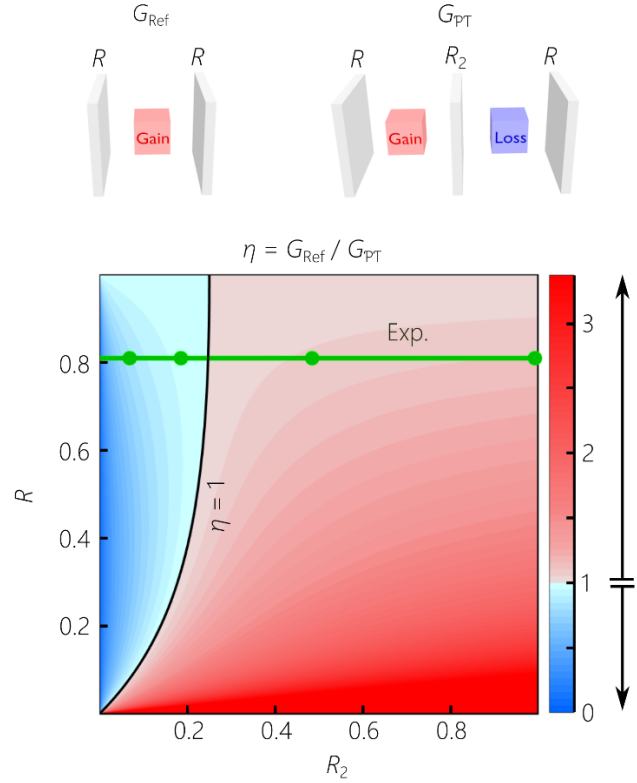

**Figure 2 | Comparison of the thresholds from a PT-symmetric cavity and an open gain-only reference cavity.** The threshold-reduction factor $\eta = G_{\text{Ref}}/G_{\text{PT}}$, where $G_{\text{Ref}}$ and $G_{\text{PT}}$ are the lasing thresholds for the reference gain cavity and the PT-symmetric cavity configurations, respectively, shown schematically at the top. We plot $\eta$ with $R$ and $R_2$, and two color palettes are used to distinguish the regime of PT-enhanced threshold where introducing the gain-balancing loss reduces the lasing threshold with respect to that of the reference gain-only cavity ($\eta > 1$, red palette), and PT-diminished threshold ($\eta < 1$, blue palette), delineated by a black curve ($\eta = 1$). The horizontal green line corresponds to the experimental value $R = 0.82$, and the circles to the experimental values of $R_2$ (see Table 1).

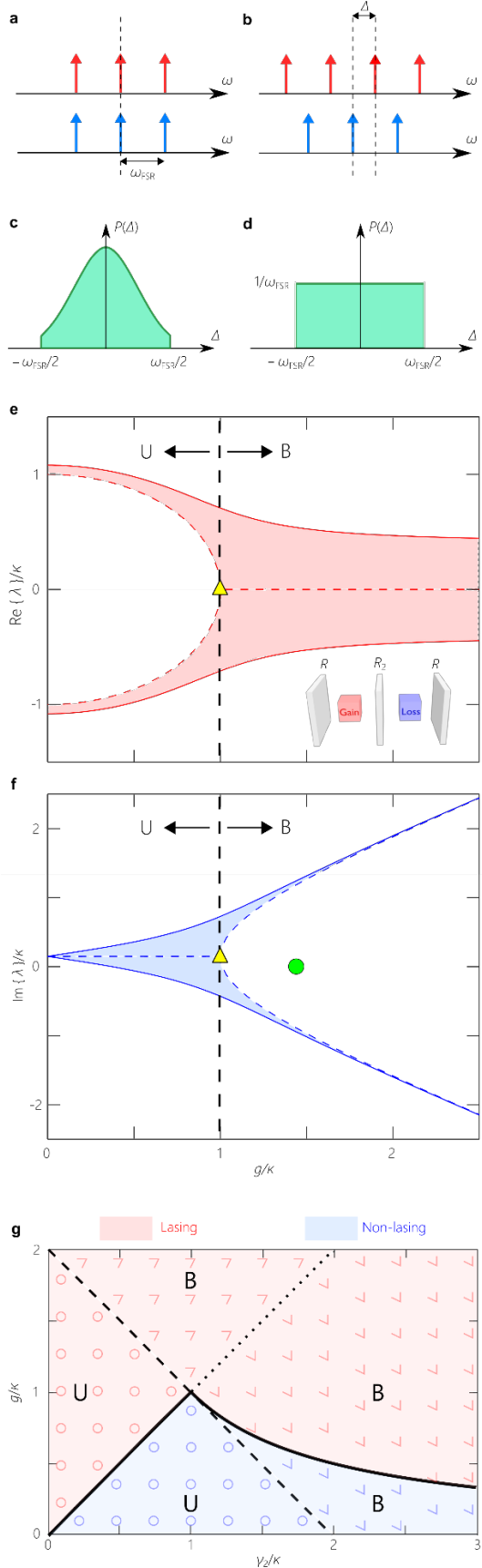

**Figure 3 | Resonance detuning and its effect on eigenvalue bifurcation. a**, In the absence of detuning, the resonance frequencies of the gain (red) and lossy (blue) sub-cavities are aligned. **b**, In the presence of detuning $\Delta$ ($-\omega_{FSR}/2 < \Delta < \omega_{FSR}/2$), the sub-cavity resonances are no longer aligned. **c-d**, Candidates for the probability distribution $P(\Delta)$ of the detuning: (**c**) a Gaussian or (**d**) a uniform distribution. **e-f**, Trajectories of (**e**) the real and (**f**) imaginary components of the eigenvalues $\lambda_{1,2}$ for a linear PT-symmetric configuration $g + \gamma_1 = \gamma_2$. Dashed curves correspond to no-detuning $\Delta = 0$, whereas the solid curves are for the case $\Delta = \omega_{FSR}/10$. The shaded regions correspond to all the intermediate detuning values. As $g$ increases, the real parts of the eigenvalues $\text{Re}\{\lambda\} = 0$ tend to coalesce whereas the corresponding imaginary parts $\text{Im}\{\lambda\}$ bifurcate. The exceptional point at zero-detuning (yellow triangle) occurs at $g = \kappa$, whereupon $\text{Re}\{\lambda\} = 0$ and $\text{Im}\{\lambda\} = \gamma_1$, thus separating the unbroken (U) and broken (B) PT-symmetric phases. The green circle corresponds to the experimental value for the lasing threshold at $R_2 = 6.8\%$ (Table 1), plotted on the axis to represent gain-clamping. Inset in (**e**) shows the PT-cavity configuration. **g**, The domains of operation of a structured cavity as dictated by the values of gain $g$ and loss $\gamma_2$ ($\gamma_1 = 0$). The dotted line $g = \gamma_2$ corresponds to the PT-symmetric condition. The dashed line $g + \gamma_2 = 2\kappa$ separates the unbroken (U, $g < 2\kappa - \gamma_2$, represented by circles) and broken (B, $g > 2\kappa - \gamma_2$, represented by edges) domains, according to Eq. (5). The lasing (red) and non-lasing (blue) regions are delineated by a solid line (the lasing threshold). In U, lasing occurs when $g > \gamma_2$, whereas lasing occurs in B when $g > \kappa^2/\gamma_2$.

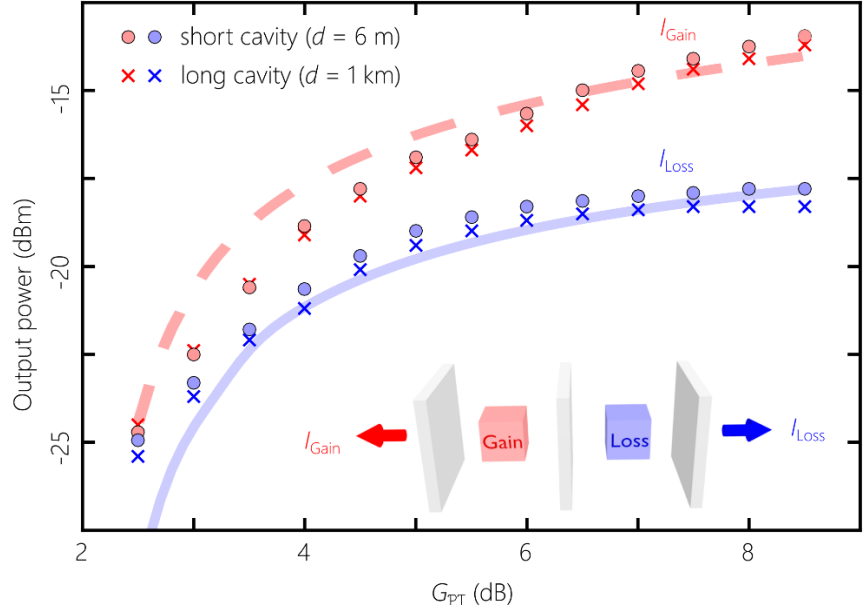

**Figure 4 | Output power-scaling from a PT-symmetric laser with gain-loss contrast.** Plot of the output power from the loss and gain laser cavity ports as the gain-loss contrast is increased while maintaining the PT-symmetric balance $G\mathcal{L} = 1$; inset shows the cavity configuration. Measured values are shown as circles and crosses for cavity configurations of total lengths $d = 6$ m and $d = 1$ km, respectively. The solid and dashed curves are simulations of the output power from loss and gain ports, respectively, obtained from the nonlinear model of the coupled fiber system in Eqs. 2-3 after making use of measured values for the model parameters and fitting the detuning $\Delta$.

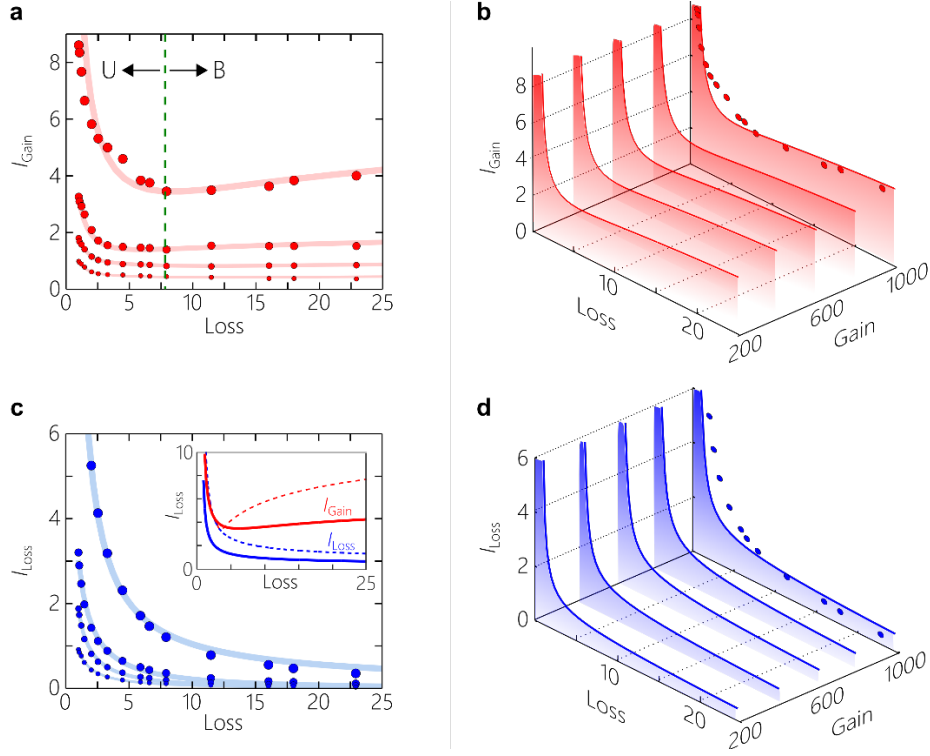

**Figure 5 | Lasing characteristics of a statistical PT-symmetric cavity around the exceptional point. a**, Measured values of the output power from the gain port $I_{\text{Gain}}$ (red circles) while varying the attenuation $\mathcal{L}$ at different gain values $G$ (30, 25, 20, and 15 dB). The solid curves are fits to guide the eye. As the loss is gradually increased at fixed gain, $I_{\text{Gain}}$ is non-monotonic. First $I_{\text{Gain}}$ decreases with $\mathcal{L}$ and goes through a minimum at the exceptional point (indicated by the vertical dashed line), and then increases with loss as the gain and loss subcavities decouple. **b**, Simulations for $I_{\text{Gain}}$ while varying the attenuation $\mathcal{L}$ at different values of $G$ using Eqs. 2-3. The red circles correspond to the data in the top-most graph in (**a**) for $G = 30$ dB. **c**, Same as (**a**) for the power from the loss port $I_{\text{Loss}}$. Inset shows theoretical plots of $I_{\text{Gain}}$ and $I_{\text{Loss}}$ at $G = 30$ dB for the statistical PT-symmetric configuration of our experiment (solid curves) and the ideal deterministic configuration (dashed curve, $\Delta = 0$) highlighting the bifurcation in output power as a consequence of PT-symmetry breaking upon passing through the system's exceptional point. **d**, Same as (**b**) for $I_{\text{Loss}}$ in lieu of $I_{\text{Gain}}$.

# Robust statistical PT-symmetric lasing in an optical fiber network

Ali K. Jahromi, Absar U. Hassan, Demetrios N. Christodoulides, and Ayman F. Abouraddy

*CREOL, The College of Optics & Photonics, University of Central Florida, Orlando, FL 32816, USA*

## Supplementary Information

# S1. Calculation of lasing thresholds using transfer matrix method

In this Section, we provide a derivation – based on the transfer matrix method – of the lasing thresholds for the structured cavity configurations compared in the main text (Fig. 2 and Table 1).

## S1.1 Transfer matrix of the cavity

We obtain the lasing threshold by extracting the poles of the transfer matrix of the full optical system. This approach assumes the poles of the transfer matrix cross the real axis of the complex frequency plane when transitioning from sub-lasing to lasing [S1,S2]. This procedure amounts to finding the conditions at which the transmission coefficient reaches a singularity and thus diverges.

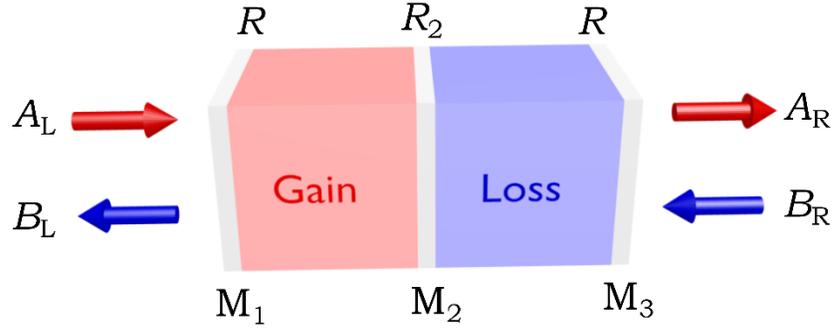

**Figure S1 | Structured lasing cavity configuration.** We depict a schematic of the structured cavity as a two-port system. The incoming amplitudes from the left and the right are $A_\mathrm{L}$ and $B_\mathrm{R}$, respectively. The outgoing amplitudes to the left and the right are $B_\mathrm{L}$ and $A_\mathrm{R}$, respectively.

Consider a structured optical cavity comprising two coupled subcavities that contain gain and loss sections, as shown schematically in Fig. S1. The mirrors used are assumed *lossless* but are *not* necessarily symmetric, so that they may be represented in general by a scattering matrix of the form [S3]

$$S = e^{i\beta} \begin{pmatrix} t & -re^{i(\beta-\alpha)} \\ re^{-i(\beta-\alpha)} & t \end{pmatrix}. \qquad (S1.1)$$

Here $t$ and $r$ are the (real) transmission and reflection coefficients, respectively, $t^2 + r^2 = 1$, $\beta$ and $\alpha$ are the transmission and reflection phases for incidence from the left, respectively, and $\beta$ and $2\beta - \alpha$ are the transmission and reflection phases for incidence from the right, respectively.

The scattering matrices act on incoming fields $\begin{pmatrix} A_L \\ B_R \end{pmatrix}$ to produce the outgoing fields $\begin{pmatrix} A_R \\ B_L \end{pmatrix}$. The reflection and transmission coefficients for the optical power are $R = r^2$ and $T = t^2$, respectively. To satisfy the PT-symmetry condition, $M_1$ and $M_3$ must have mirror-symmetry with respect to the center of the structure. If we take $M_1$ to be represented by the $S$-matrix in Eq. S1.1, then this symmetry requirement dictates that the $S$-matrix for $M_3$ has the form:

$$S = e^{i\beta} \begin{pmatrix} t & re^{-i(\beta-\alpha)} \\ -re^{i(\beta-\alpha)} & t \end{pmatrix}. \tag{S1.2}$$

Furthermore, the mirror $M_2$ must be symmetric, which adds the constraint to its $S$-matrix that $\beta = \alpha - \pi/2$, such that its S-matrix has the simple form:

$$S = e^{i\beta} \begin{pmatrix} t & ir \\ ir & t \end{pmatrix}. \tag{S1.3}$$

Exploiting the above-described scattering matrices, we obtain the corresponding transfer matrices that act on the forward and backward fields $\begin{pmatrix} A_R \\ B_R \end{pmatrix}$ on the right to produce the corresponding fields $\begin{pmatrix} A_L \\ B_L \end{pmatrix}$ on the left. The transfer matrices $M_1$, $M_2$, and $M_3$ representing the mirrors are given by

$$M_1 = \frac{1}{t}\begin{pmatrix} e^{-i\beta} & re^{i(\beta-\alpha)} \\ re^{-i(\beta-\alpha)} & e^{i\beta} \end{pmatrix}, M_2 = \frac{1}{t_2}\begin{pmatrix} e^{-i\beta_2} & -ir_2 \\ ir_2 & e^{i\beta_2} \end{pmatrix}, M_3 = \frac{1}{t}\begin{pmatrix} e^{-i\beta} & -re^{-i(\beta-\alpha)} \\ -re^{i(\beta-\alpha)} & e^{i\beta} \end{pmatrix}.$$

$$\tag{S1.4}$$

The parameters in these transfer matrices are not subject to statistical fluctuations, except perhaps the phases if the mirrors are implemented by extended fiber Bragg gratings (FBGs).

The transfer matrices for the intervening gain and loss layers are given by

$$M_G = \begin{pmatrix} \frac{1}{\sqrt{G}}e^{-i\varphi_g} & 0 \\ 0 & \sqrt{G}e^{i\varphi_g} \end{pmatrix}, M_{\mathcal{L}} = \begin{pmatrix} \frac{1}{\sqrt{\mathcal{L}}}e^{-i\varphi_\ell} & 0 \\ 0 & \sqrt{\mathcal{L}}e^{i\varphi_\ell} \end{pmatrix}, \tag{S1.5}$$

where $G$ and $\mathcal{L}$ are single-pass amplification and attenuation factors for the power traversing the gain and loss layers, respectively, and $\varphi_g$ and $\varphi_\ell$ are the single-pass propagation phases in the gain or loss layers, respectively. In the exact PT-symmetric configuration, $G\mathcal{L} = 1$ and $\varphi_g = \varphi_\ell$. However, statistical fluctuations in a macroscopic cavity precludes the realization of the latter condition.

The total transfer matrix for the whole cavity is thus

$$M = M_1 \, M_G \, M_2 \, M_\mathcal{L} \, M_3 = \begin{pmatrix} m_{11} & m_{12} \\ m_{21} & m_{22} \end{pmatrix}. \tag{S1.6}$$

The cavity transmission, which shares the same poles as the field transmission coefficient, is $T_L = 1/|m_{11}|^2 = 1/M_{11}$, where

$$M_{11} = \frac{1}{(1-R)^2(1-R_2)} \left\{ \frac{1}{G\mathcal{L}} + \frac{G}{\mathcal{L}} RR_2 + \frac{\mathcal{L}}{G} RR_2 + G\mathcal{L}R^2 + 2\sqrt{RR_2}\left(\frac{1}{\mathcal{L}} + \mathcal{L}R\right)\cos\varphi_G + 2\sqrt{RR_2}\left(\frac{1}{G} + GR\right)\cos\varphi_\mathcal{L} + 2R[\cos(\varphi_G + \varphi_\mathcal{L}) + R_2\cos(\varphi_G - \varphi_\mathcal{L})] \right\}. \tag{S1.7}$$

Therefore the cavity transmission is determined by (1) the mirror reflectivities $R$ and $R_2$, (2) the gain $G$ and the loss $\mathcal{L}$, and (3) two phases $\varphi_G$ and $\varphi_\mathcal{L}$:

$$\varphi_G = 2\varphi_g + 2\beta - \alpha + \beta_2 + \pi/2,$$

$$\varphi_\mathcal{L} = 2\varphi_\ell + 2\beta - \alpha + \beta_2 + \pi/2. \tag{S1.8}$$

## S1.2 Lasing thresholds

Since the cavity length $L_\text{cavity}$ is very large with respect to the optical wavelength $\lambda$, $L_\text{cavity} \gg \lambda$, the single-pass phases may not be deterministic due to minute thermal or mechanical fluctuations in the optical fibers used. We thus assume that the cavity single-pass phases are random each with a uniform probability distribution in the interval from 0 to $2\pi$. Consequently, lasing may occur at any wavelength within the gain and loss bandwidth. To obtain the lasing threshold, we hold fixed the physical parameters related to the optical loss and mirror reflectivities, while varying the gain. After obtaining the lasing gain threshold, we can find the special case for the PT-symmetric configuration by choosing the attenuation to satisfy $G\mathcal{L} = 1$.

At any value of gain, lasing is initiated whenever the cavity phases produce a zero in $M_{11}$ (a transmission pole). We therefore minimize $M_{11}$ with respect to the phases, and then extract the zeros to find the gain corresponding to the lasing threshold. We thus first set $\frac{\partial M_{11}}{\partial \varphi_G} = \frac{\partial M_{11}}{\partial \varphi_\mathcal{L}} = 0$, from which we find a minimal value of $M_{11}$ when $\varphi_G = 0$ and $\varphi_\mathcal{L} = \pi$,

$$M_{11}|_\text{min} = M_{11}(\varphi_G = 0, \varphi_\mathcal{L} = \pi) = \frac{1}{T^2 T_2} \frac{1}{G\mathcal{L}} \left[ G(\mathcal{L}R + \sqrt{RR_2}) - (1 + \mathcal{L}\sqrt{RR_2}) \right]^2. \tag{S1.9}$$

From this minimal value of $M_{11}$, we obtain the zero at a threshold gain $G_\text{th}$ given by

$$G_\text{th} = \frac{1}{\sqrt{R}} \frac{1 + \mathcal{L}\sqrt{RR_2}}{\mathcal{L}\sqrt{R} + \sqrt{R_2}}. \tag{S1.10}$$

This general expression for the gain threshold is the minimum required gain for lasing to occur, while all other parameters are fixed except for the phases that are assumed to vary randomly.

We can now obtain expressions for the gain threshold when special structures are of interest. First, for a pseudo-PT-symmetric structure with $G\mathcal{L} = 1$, the following threshold holds for lasing:

$$G_{\text{PT}} = \frac{1-R}{2\sqrt{RR_2}} + \sqrt{1 + \left(\frac{1-R}{2\sqrt{RR_2}}\right)^2}, \tag{S1.11}$$

which was used in the main text (Eq. 1). Second, when we eliminate the attenuation in the loss subcavity $\mathcal{L} = 1$, the gain threshold $G_0$ has the form

$$G_0 = \frac{1}{\sqrt{R}} \frac{1+\sqrt{RR_2}}{\sqrt{R}+\sqrt{R_2}}. \tag{S1.12}$$

Finally, when the lossy subcavity is altogether open, corresponding to $\mathcal{L} = 0$, the gain threshold $G_{\text{open}}$ has the usual form

$$G_{\text{open}} = \frac{1}{\sqrt{RR_2}}. \tag{S1.13}$$

## S2. Non-Hermitian temporal coupled-mode equations

We now introduce the temporal coupled-mode equations used as a basis for the analysis presented in the main text. We consider an optical-fiber-based laser cavity, comprising two coupled sub-cavities (one including net gain and the other net loss) of length $L$ each and group velocity $v_g$. We introduce time-dependent quantities to characterize the sub-cavities: a linear background loss per second $\gamma_1$ for the field in the gain sub-cavity (which only depends on the side mirror M$_1$ of reflectivity $R$) and a corresponding quantity $\gamma_2$ for the loss sub-cavity (which depends on the side mirror M$_3$ of reflectivity $R$ in addition to the imposed attenuation). Furthermore, we define a temporal small-signal gain $g$ for the gain sub-cavity, which is dictated by the SOA amplification. These temporal losses and the small-signal gain are given by

$$\gamma_1 = \frac{v_g}{2L} \ln\left(\frac{1}{\sqrt{R}}\right), \tag{S2.1}$$

$$\gamma_2 = \frac{v_g}{2L} \left[\ln\left(\frac{1}{\Gamma\sqrt{R}}\right)\right], \tag{S2.2}$$

$$g = \frac{v_g}{2L} \ln(G), \tag{S2.3}$$

where $G$ and $\Gamma$ are the single-pass intensity amplification and attenuation factors that are set by the SOA and the VOA, respectively.

Based on these considerations, we define two temporal coupled-mode equations for the mean-field amplitudes $\tilde{a}$ and $\tilde{b}$ in the amplifying and attenuating fiber sub-cavities, respectively, which adequately capture all the essential features of the optical field dynamics in the coupled fiber-cavity system:

$$\frac{d\tilde{a}}{dt} = -\gamma_1 \tilde{a} + i\frac{\Delta}{2}\tilde{a} + \left(\frac{g}{1+|\tilde{a}|^2/I_s}\right)\tilde{a} + i\kappa\tilde{b}, \tag{S2.4}$$

$$\frac{d\tilde{b}}{dt} = -\gamma_2 \tilde{b} - i\frac{\Delta}{2}\tilde{b} + i\kappa\tilde{a}. \tag{S2.5}$$

We have introduced into this model the nonlinear gain-saturation, the loss mechanisms, and the coupling between the sub-cavities, where $I_s$ is the gain-saturation intensity and $\Delta$ is used to phenomenologically introduce detuning between the resonances of the sub-cavities. For simplicity, we normalize the field amplitudes with respect to the saturation value by introducing the scaled amplitudes $(\tilde{a}, \tilde{b}) = \sqrt{I_s}(a, b)$, whereupon we obtain the coupled-mode equations:

$$\frac{da}{dt} = -\gamma_1 a + i\frac{\Delta}{2}a + \left(\frac{g}{1+|a|^2}\right)a + i\kappa b, \tag{S2.6}$$

$$\frac{db}{dt} = -\gamma_2 b - i\frac{\Delta}{2}b + i\kappa a. \tag{S2.7}$$

## S3. Extended Lamb model for coupled cavities

In traditional configurations of coupled cavities, such as evanescently coupled micro-ring resonators [S4], one considers *time-averaged* energy amplitudes in each cavity. The temporal coupling between the cavities is dependent on the effective interaction region between them. In our setup for coupled fiber cavities, their interaction is mediated by the middle mirror $M_2$, which is a lumped element (Fig. 1 in the main text and Fig. S2). The usual approach of extracting a distributed effective coupling is thus no longer valid. A more pertinent method was employed by W. E. Lamb [S5,S6] in treating coupled-laser configurations, in which the central mirror coupling two sub-cavities (Fig. S2a) was modelled as a "bump" of width $w$ in the permittivity distribution in space along the cavity axis $z$ (Fig. S2b). The permittivity of the so-called "bump" is defined as:

$$\varepsilon(z) = \varepsilon_0\left(n_0^2 + \rho\,\delta(z)\right), \tag{S3.1}$$

where $n_0$ is the refractive index of the background medium. Lamb's well-known model thus connects the sought-after coupling coefficient $\kappa$ (Eqs. 2-3 in the main text and Sections S2 and S4-S6 below) to the "bump" parameter $\rho$:

$$\kappa = \frac{v_g}{k\rho L}, \tag{S3.2}$$

according to Eq. 18 of reference [S4]; here $L$ is the length of a single sub-cavity, $k = n_0\, 2\pi/\lambda$, and $v_g$ is the group velocity of light in the fibers.

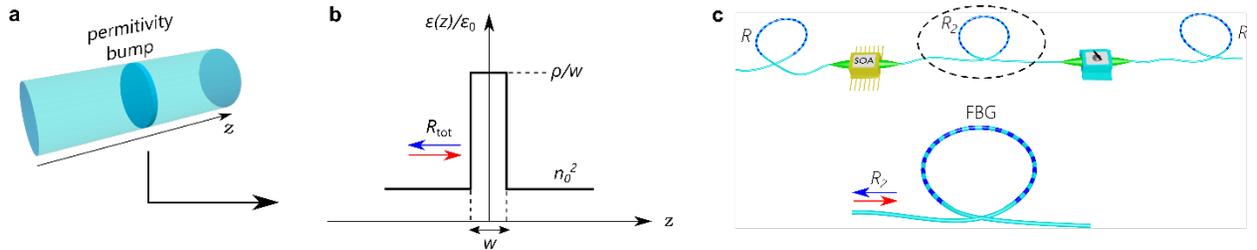

**Figure S2 | Extended Lamb model to extract the coupling coefficient between the sub-cavities in the PT-symmetric cavity. a**, Cavity used in the Lamb model in which a discontinuity (a permittivity "bump") is introduced in a laser cavity. **b**, Model for the permittivity "bump" of width $w$. **c**, The PT-symmetric laser configuration used in our experiments. The inset in the FBG used as $M_2$, which corresponds to the permittivity "bump" in Lamb's model.

In our experiment, the bump represents the mirror $M_2$ of reflectivity $R_2$. By establishing a relationship between the bump parameter $\rho$ and the reflectivity $R_2$, we can then express $\kappa$ in terms of $R_2$. This connection enables us to simulate the dynamics of our structure via coupled-mode theory after employing the proper coupling coefficient.

According to the Fig. S2b, the discontinuity in the refractive index ($n = \sqrt{\rho/w}$) results in Fresnel reflection at each interface, $R_{\text{int}}$:

$$R_{\text{int}} = \left(\frac{n-n_0}{n+n_0}\right)^2 = \left(\frac{\sqrt{\rho/w}-n_0}{\sqrt{\rho/w}+n_0}\right)^2 \cong 1 - 4n_0\sqrt{w/\rho}, \tag{S3.3}$$

Here the subscript 'int' indicates 'interface' and the approximation makes use of the fact that $n = \sqrt{\rho/w}$ grows in the limit $w \to 0$. The two interfaces of the "bump" define a Fabry-Pérot resonator with a reflection of $R_{\text{tot}}$ (Fig. S2b):

$$R_{\text{tot}} = 1 - \frac{1}{1+\left[\frac{2\sqrt{R_{\text{int}}}}{1-R_{\text{int}}}\sin\left(\frac{2\pi}{\lambda}nw\right)\right]^2}, \tag{S3.4}$$

Taking the limit $w \to 0$ leads to the approximation $\sin(\pi nw/\lambda) = \sin(\pi\sqrt{\rho w}/\lambda) \cong \pi\sqrt{\rho w}/\lambda$, and substituting $R_{\text{int}}$ from Eq. S2.3 into Eq. S2.4, so that the total reflection of the "bump" is:

$$R_{\text{tot}} \cong \frac{1}{1+\left(\frac{n_0\lambda}{\pi\rho}\right)^2}. \tag{S3.5}$$

We associate the reflectivity of the mirror $R_2$ with that of the 'bump'-based model $R_{\text{tot}}$, $R_{\text{tot}} \to R_2$. Therefore $\rho$ can now be expressed in terms of $R_2$,

$$\rho = \frac{n_0\lambda}{\pi}\sqrt{\frac{R_2}{1-R_2}}. \tag{S3.6}$$

According to Lamb's model, the coupling rate is related to $\rho$ (and thus to $R_2$) via

$$\kappa = \frac{v_g}{k\rho L} = \frac{v_g}{2n_0^2 L}\sqrt{\frac{1-R_2}{R_2}}, \tag{S3.7}$$

Coupled-mode theory simulations that make use of $\kappa$ defined in this last expression do not quite agree with our measurements. The reason is that the Lamb model assumes that the out-coupling windows $M_1$ and $M_3$ are perfect, which is not the case in our configuration. In another seminal paper, W. E. Lamb considered the case of partially transmitting windows that radiate into free space [S6]. The analysis revealed that for windows with a high enough reflectivity (as is the case in our experiment), only a small leakage and a change in the cavity free-spectral-range (FSR) are effectively introduced – whereas the modal profiles are left unchanged. The change in the FSR means that the effective length of the cavity is now dependent upon the out-coupling mirror reflectivity. We model this effect in the expression for the coupling between the two sub-cavities of the PT-laser as follows:

$$\kappa = \frac{v_g}{2n_0^2 L}(1-R)\sqrt{\frac{1-R_2}{R_2}}. \tag{S3.8}$$

This is the definition of $\kappa$ used in our coupled-mode theory simulations throughout. The value of $\kappa$ obtained as such, when used to find the lasing threshold $g_{th}$ based on Eqs. (2-3) of the main text, agrees well with the threshold found based on a transfer matrix analysis as provided in section S1.2. When one considers the scenario of $G\Gamma = 1$, $\gamma_2$ is always greater than $g$, in fact $\gamma_2 = \gamma_1 + g$ – see Eqs. S2.2 and S2.3. Now the condition for lasing in an unbroken mode is given by, $g > (\gamma_1 + \gamma_2)$ which can never be achieved (consider the imaginary part of Eq. S6.3). However, the lasing threshold in the broken mode might be surpassed. In this case when one enters the broken symmetry regime where the square root in Eq. S6.3 turns complex, the threshold can be obtained from the requirement: $\gamma_1 - \kappa\sqrt{\left(\frac{g}{\kappa}\right)^2 - 1} < 0$. This then leads to,

$$g_{th} = \sqrt{\gamma_1^2 + \kappa^2}. \qquad (S3.9)$$

Translating this into a single-pass gain value using Eq. S2.3, one obtains,

$$G_{th} = \exp\left\{\sqrt{\left[\ln\left(\frac{1}{\sqrt{R}}\right)\right]^2 + \left[\frac{(1-R)}{n_0^2}\sqrt{\frac{1-R_2}{R_2}}\right]^2}\right\}. \qquad (S3.10)$$

The value of this threshold is compared to $G_{PT}$ (according to Eq. S1.11) for various values of $R_2$ in the figure below. A good agreement between the two is apparent.

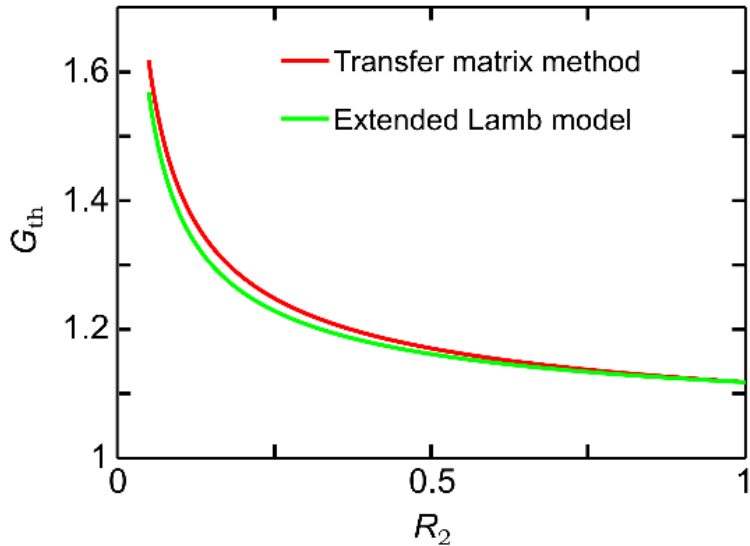

**Figure S3 | Comparison of the lasing thresholds for a PT-symmetric lasing cavity obtained from the transfer matrix method and the extended Lamb model.** The lasing thresholds are plotted as a function of $R_2$ with both out-coupling mirror reflectivities assumed to be $R = 80\%$.

## S4. Temporal coupled-mode solutions in absence of detuning

We first obtain solutions to the coupled-mode equations (Eqs. S2.6-7) in absence of detuning, $\Delta = 0$ to obtain the coupled-mode equations:

$$\frac{da}{dt} = -\gamma_1 a + \left(\frac{g}{1+|a|^2}\right)a + i\kappa b, \tag{S4.1}$$

$$\frac{db}{dt} = -\gamma_2 b + i\kappa a. \tag{S4.2}$$

We seek a harmonic solution of the form $\begin{pmatrix} a(t) \\ b(t) \end{pmatrix} = \begin{pmatrix} a_0 \\ b_0 \end{pmatrix} e^{i\lambda t}$, where $\lambda$ is real and $\begin{pmatrix} a_0 \\ b_0 \end{pmatrix}$ is a constant vector. When $|a_0| = |b_0|$, we have a so-called 'unbroken' PT-symmetric phase, otherwise, when $|a_0| \neq |b_0|$ we obtain a 'broken' phase. By substituting this solution into Eq. S4.1-2 we obtain

$$i\lambda + \gamma_1 - g_s = i\kappa \frac{b_0}{a_0}, \tag{S4.3}$$

$$i\lambda + \gamma_2 = i\kappa \frac{a_0}{b_0}, \tag{S4.4}$$

where, as before, $g_s = \frac{g}{1+|a_0|^2}$ for the saturated gain, and we assume that $|a_0|, |b_0| \neq 0$. Multiplying Eq. S4.3 and Eq. S4.4, we obtain a quadratic equation in $\lambda$,

$$\lambda^2 - i(\gamma_1 + \gamma_2 - g_s)\lambda - \kappa^2 - \gamma_2(\gamma_1 - g_s) = 0. \tag{S4.5}$$

This quadratic equation has a real solution for $\lambda$ only when $\gamma_1 + \gamma_2 = g_s$, whereupon

$$|a_0|^2 = \frac{g}{\gamma_1+\gamma_2} - 1, \tag{S4.6}$$

and the solutions for $\lambda$ take the simple form:

$$\lambda^2 = \kappa^2 - \gamma_2^2 \rightarrow \lambda_{1,2} = \pm\kappa\sqrt{1 - \left(\frac{\gamma_2}{\kappa}\right)^2}, \tag{S4.7}$$

which applies when $\gamma_2 \leq \kappa$. By defining $\gamma_2 = \kappa \sin\theta$, we obtain a concise expression for the eigenvalues: $\lambda_{1,2} = \pm\kappa\cos\theta$.

To obtain the complex ratio between $b_0$ and $a_0$, we define $b_0 = a_0 \alpha e^{i\phi}$ and subtract Eq. S4.4 from S4.3,

$$-2\gamma_2 = i\kappa\left(\alpha e^{i\phi} - \frac{1}{\alpha e^{i\phi}}\right), \tag{S4.8}$$

which leads to

$$2i\sin\theta = \alpha\cos\phi + i\alpha\sin\phi - \frac{1}{\alpha}\cos\phi + i\frac{1}{\alpha}\sin\phi. \tag{S4.9}$$

Comparing the real parts of Eq. S4.9 yields

$$\alpha\cos\phi = \frac{1}{\alpha}\cos\phi \rightarrow \alpha = \pm 1. \tag{S4.10}$$

Now, comparing the imaginary parts of Eq. S4.9 we have

$$2 \sin \theta = \alpha \sin \phi + \frac{1}{\alpha} \sin \phi. \tag{S4.11}$$

We now have a pair of conditions $\alpha = 1 \rightarrow \phi = \theta$ and $\alpha = -1 \rightarrow \phi = -\theta$. Hence, the steady-state solution is given by

$$\begin{pmatrix} a \\ b \end{pmatrix} = \sqrt{\frac{g}{\gamma_1+\gamma_2} - 1} \begin{pmatrix} 1 \\ \pm e^{\pm i\theta} \end{pmatrix} e^{\pm i(\kappa \cos \theta)t}, \tag{S4.12}$$

where $\theta$ is obtained from $\sin \theta = \gamma_2/\kappa$. We therefore obtain a pure 'unbroken' PT-symmetric phase when there is no detuning, that is valid when $\gamma_2 \leq \kappa$ and $g > \gamma_1 + \gamma_2$. The latter restriction is inferred from Eq. S4.6.

On the other hand, if $\gamma_2 > \kappa$, we have stationary steady-state solutions $\lambda = 0$ in Eq. S4.3-4,

$$0 = -\gamma_1 + \frac{g}{1+|a_0|^2} + i\kappa \frac{b_0}{a_0}, \tag{S4.13}$$

$$0 = -\gamma_2 + i\kappa \frac{a_0}{b_0}. \tag{S4.14}$$

Equation S4.14 implies that $b_0 = i\frac{\kappa}{\gamma_2} a_0$, which we substitute in Eq. S4.13 to obtain

$$|a_0|^2 = \frac{g}{\gamma_1+\kappa^2/\gamma_2} - 1. \tag{S4.15}$$

The solution within this regime is thus given by,

$$\begin{pmatrix} a \\ b \end{pmatrix} = \sqrt{\frac{g}{\gamma_1+\kappa^2/\gamma_2} - 1} \begin{pmatrix} 1 \\ i\kappa/\gamma_2 \end{pmatrix}, \tag{S4.16}$$

which corresponds to a PT-symmetry-broken phase solution that is valid when $\gamma_2 > \kappa$ and $g > \gamma_1 + \kappa^2/\gamma_2$. The last condition is a consequence of having a positive absolute value $|a_0|^2$ in Eq. S4.15.

In summary, steady-state solutions of the coupled-mode equations Eqs. S4.1-2 have been found in two different regimes:

(I) Unbroken: when $g > \gamma_1 + \gamma_2$ and $\geq \gamma_2 \rightarrow \begin{pmatrix} a \\ b \end{pmatrix} = \sqrt{\frac{g}{\gamma_1+\gamma_2} - 1} \begin{pmatrix} 1 \\ \pm e^{\pm i\theta} \end{pmatrix} e^{\pm i(\kappa \cos \theta)t}$;

(II) Broken: when $g > \gamma_1 + \kappa^2/\gamma_2$ and $< \gamma_2 \rightarrow \begin{pmatrix} a \\ b \end{pmatrix} = \sqrt{\frac{g}{\gamma_1+\kappa^2/\gamma_2} - 1} \begin{pmatrix} 1 \\ i\kappa/\gamma_2 \end{pmatrix}$.

## S5. Proof for the absence of a pure unbroken-symmetry mode in presence of detuning

In the presence of detuning $\Delta \neq 0$, it can be shown that a formal unbroken PT-symmetric phase (that is, one with equal intensities of harmonic fields at the two output ports $|a_0| = |b_0|$) cannot be realized. This fact can be deduced from Eqs. S2.6-7 where a harmonic solution $\binom{a}{b} = \binom{a_0}{b_0} e^{i\lambda t}$ with $\lambda \in \mathbf{R}$ results in the following equations:

$$\lambda = i\gamma_1 + \frac{\Delta}{2} - ig_s + \kappa \frac{b_0}{a_0}, \tag{S5.1}$$

$$\lambda = i\gamma_2 - \frac{\Delta}{2} + \kappa \frac{a_0}{b_0}, \tag{S5.2}$$

where $g_s = g/(1 + |a_0|^2)$. From Eq. S5.2, we obtain the ratio between $a_0$ and $b_0$,

$$b_0 = i \frac{\kappa}{\gamma_2 + i\delta} a_0, \tag{S5.3}$$

where $\delta = \lambda + \frac{\Delta}{2}$. Substituting in Eq. S5.1, and assuming that $a_0 \neq 0$, we obtain

$$\left\{\gamma_1 - g_s + \frac{\kappa^2 \gamma_2}{\gamma_2^2 + \delta^2}\right\} + i\left\{\delta - \Delta - \frac{\kappa^2 \delta}{\gamma_2^2 + \delta^2}\right\} = 0. \tag{S5.4}$$

The imaginary part of this equation implies that

$$\frac{\gamma_2^2 + \delta^2}{\kappa^2} = \frac{\delta}{\delta - \Delta}. \tag{S5.5}$$

Achieving the unbroken PT-symmetric phase requires $|b_0| = |a_0|$, which necessitates (via Eq. S5.3) that

$$\frac{\kappa^2}{\gamma_2^2 + \delta^2} = 1 \quad \rightarrow \quad \frac{\delta - \Delta}{\delta} = 1 \tag{S5.6}$$

Such a result is only possible in one of two scenarios: (1) $\Delta = 0$, that is, no detuning; or (2) $\delta \rightarrow \infty$. Hence for a *non-zero* detuning, the exact unbroken PT phase in the two coupled sub-cavities *never* appears.

## S6. Linear analysis in the presence of detuning

Before the onset of lasing in the coupled-cavity structure, a linear analysis of the field dynamics can be carried out because the intensities are small. Under these conditions, the dynamical equations can be cast in the form:

$$\frac{da}{dt} = -\gamma_1 a + i\frac{\Delta}{2}a + ga + i\kappa b, \tag{S6.1}$$

$$\frac{db}{dt} = -\gamma_2 b - i\frac{\Delta}{2}b + i\kappa a. \tag{S6.2}$$

Detuning leads to an avoided *eigenvalue-coalescence*, as we proceed to show. First, we consider the harmonic ansatz $\begin{pmatrix} a(t) \\ b(t) \end{pmatrix} = \begin{pmatrix} a_0 \\ b_0 \end{pmatrix} e^{i\lambda t}$ in absence of detuning, i.e. $\Delta = 0$, to obtain:

$$\lambda_{1,2} = -\frac{i}{2}(g - \gamma_1 - \gamma_2) \pm \kappa\sqrt{1 - \left(\frac{g - \gamma_1 + \gamma_2}{2\kappa}\right)^2}. \tag{S6.3}$$

whereupon a second order bifurcation takes place in the imaginary parts of the eigenvalue $\lambda$ at $g = 2\kappa + \gamma_1 - \gamma_2$ by gradually increasing the gain from a small value. The system is said to enter a PT-symmetry broken regime when the amount of gain increases beyond this point. That is, the *unbroken* PT-symmetry regime occurs when $g < \gamma_1 - \gamma_2 + 2\kappa$ (the eigenvalues have not undergone a bifurcation) and the *broken* PT-symmetry regime when $g > \gamma_1 - \gamma_2 + 2\kappa$ (the eigenvalues have bifurcated). This transition point between broken and unbroken PT-symmetry regimes is better known as an 'exceptional point' (EP) [S7-S9]. At this point, the two eigenvalues and associated eigenvectors $(a_0, b_0)^T$ coalesce:

$$\lambda_{1,2} = i(\gamma_2 - \kappa), \quad |1,2\rangle = \begin{pmatrix} 1 \\ i \end{pmatrix}. \tag{S6.4}$$

In general, lasing occurs once the sign of the imaginary part of one of the eigenvalues $\lambda_{1,2}$ becomes negative, $\text{Im}\{\lambda\} < 0$, per our harmonic ansatz. Because the imaginary parts of the eigenvalues drop monotonically with $g$ prior to the EP (Eq. S6.3), the expression for the eigenvalues at the EP (Eq. S6.4) dictates whether lasing is initiated in the unbroken or broken PT-symmetry regimes. If $\text{Im}\{\lambda\} < 0$ at the EP because $\kappa > \gamma_2$, this indicates that lasing has already started before the EP was reached. In other words, lasing initiates here in the unbroken PT-symmetry regime. On the other hand, if $\text{Im}\{\lambda\} > 0$ at the EP because $\kappa < \gamma_2$, then lasing has not started when the EP is reached. Therefore, lasing initiates after the EP in the broken PT-symmetry regime. These two scenarios of lasing occurring in the unbroken and broke PT-symmetric phases are depicted in Fig. S4.

This picture of pure unbroken- and broken-symmetry regimes is true only for a zero-detuned system ($\Delta = 0$). Once detuning is introduced ($\Delta \neq 0$), the eigenvalues do not coalesce, as shown in Fig. S5. In other words, an exact EP does not occur. Nevertheless, two regimes are still identifiable, one in which the eigenvalues approach each other (unbroken PT-symmetry) and the other where they diverge (broken PT-symmetry). In the presence of detuning $\Delta \neq 0$, the eigenvalues are

$$\lambda_{1,2} = -\frac{i}{2}(g - \gamma_1 - \gamma_2) \pm \kappa\sqrt{1 - \left(\frac{g - \gamma_1 + \gamma_2 + i\Delta}{2\kappa}\right)^2}. \tag{S6.5}$$

We can infer from the expression under the square root in Eq. S6.5 that $\lambda_1 \neq \lambda_2$ when $\Delta \neq 0$ because the parameters $g$, $\gamma$, and $\kappa$ are all real. Using $g - \gamma_1 + \gamma_2 + i\Delta = 2\kappa \sin\theta$, where $\theta$ is now a complex angle, the eigenvectors associated with these eigenvalues are

$$|1\rangle = \begin{pmatrix} 1 \\ e^{i\theta} \end{pmatrix}, |2\rangle = \begin{pmatrix} 1 \\ -e^{-i\theta} \end{pmatrix}. \tag{S6.6}$$

Note that the EP in the zero-detuning scenario corresponds to $\theta = \pi/2$.

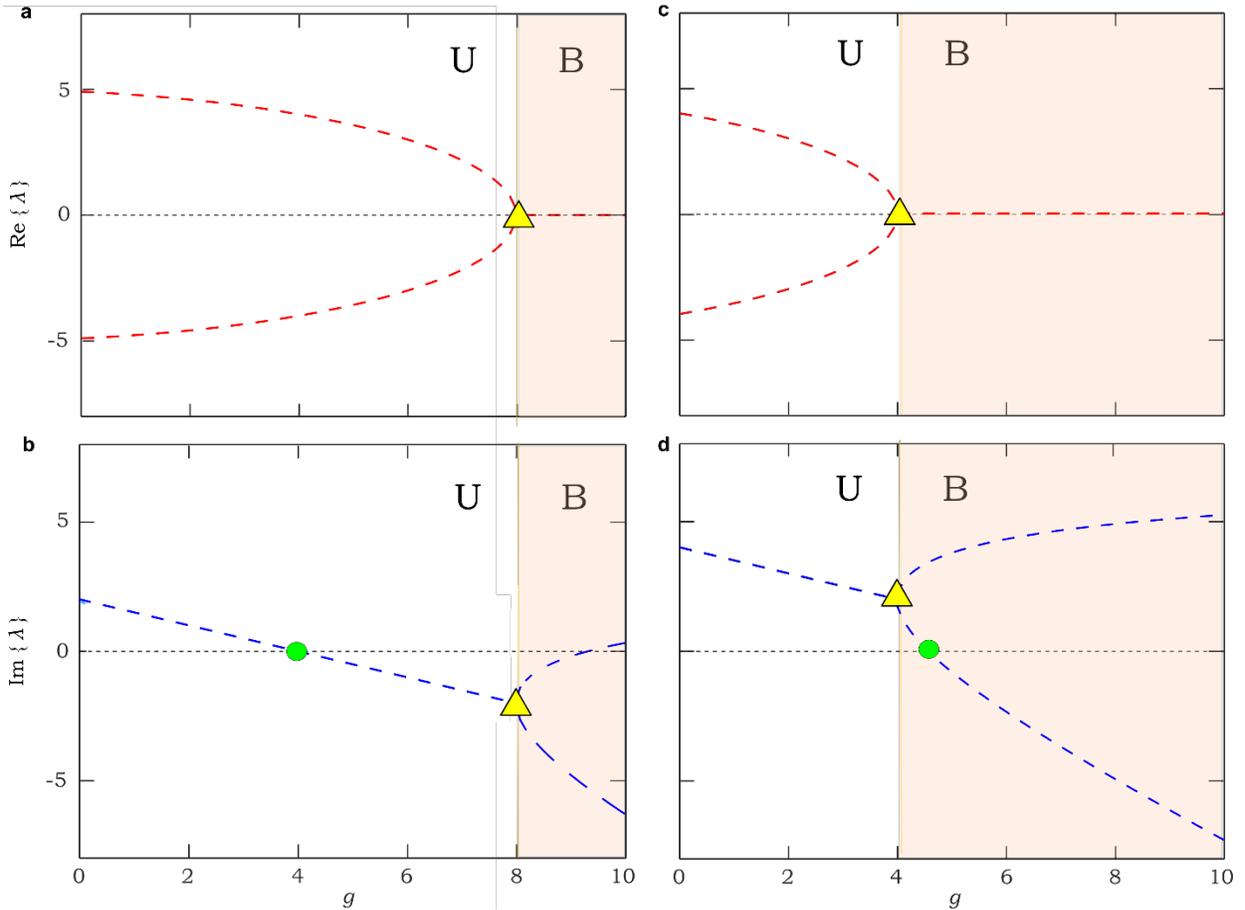

**Figure S4 | Eigenvalue dynamics, exceptional points, and lasing in two coupled sub-cavities at zero-detuning. a-d**, Trajectories of the two eigenvalues $\lambda_1$ and $\lambda_2$ of the harmonic solutions to the temporal coupled-mode theory treatment in absence of detuning. The real parts are depicted in the top row and the imaginary parts in the bottom row. The light and dark backgrounds identify the unbroken ('U') and broken ('B') PT-symmetric regimes, respectively. EP is depicted as a yellow triangle (where $g = 2\kappa + \gamma_1 - \gamma_2$) and the green circle indicates the point where lasing occurs. In all cases, we use the parameter values $\gamma_1 = 1$ and $\kappa = 5$. **a-b**, When the loss $\gamma_2 = 3$ is less than the coupling strength $\kappa$, lasing occurs in the unbroken regime. **c-d**, When the loss $\gamma_2 = 7$ is larger than the coupling strength $\kappa$, lasing is initiated in the broken regime instead.

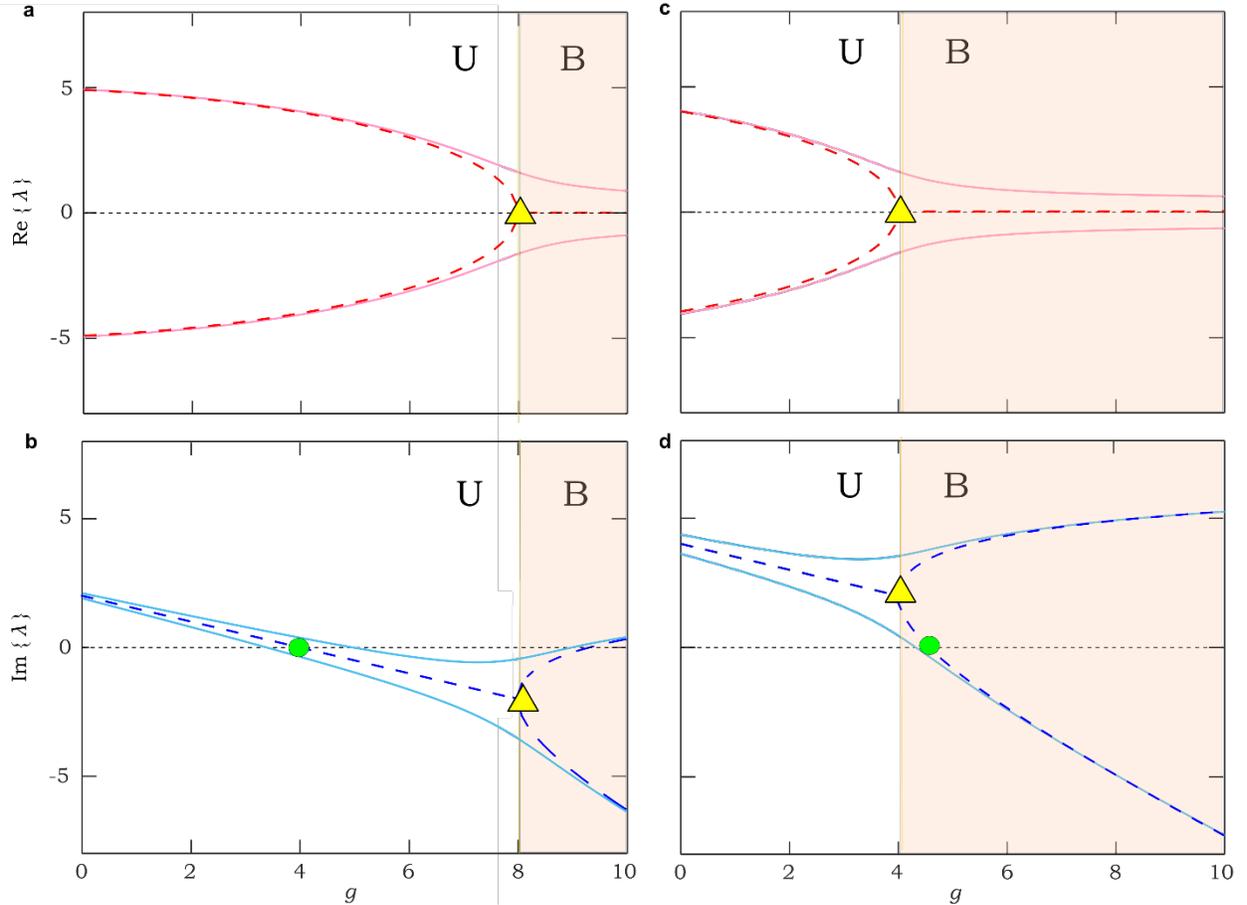

**Figure S5 | Eigenvalue dynamics, exceptional points, and lasing in two coupled sub-cavities in presence of detuning. a-d**, Trajectories of the two eigenvalues $\lambda_1$ and $\lambda_2$ of the harmonic solutions to the temporal coupled-mode theory treatment for $\Delta \neq 0$. The real parts are depicted in the top row and the imaginary parts in the bottom row. Dashed and solid curves correspond to the eigenvalues in absence and presence of detuning, respectively. The light and dark backgrounds identify the unbroken ('U') and broken ('B') PT-symmetric regimes, respectively. EP is depicted as a yellow circle and the green circle indicates the point where lasing occurs in the corresponding zero-detuning case In all cases, we use the parameter values $\gamma_1 = 1$, $\kappa = 5$, and $D = 1$. **a-b**, When the loss $\gamma_2 = 3$ is less than the coupling strength $\kappa$, lasing occurs in the unbroken regime. **c-d**, When the loss $\gamma_2 = 7$ is larger than the coupling strength $\kappa$, lasing is initiated in the broken regime instead.

The real parts of the two eigenvalues are plotted in Fig. S4 and Fig. S5 against gain $g$ in the case where $\gamma_2$ is kept fixed at $\gamma_2 = 3$, the coupling is $\kappa = 5$, and the loss is $\gamma_1 = 1$. The two eigenvalues are initially well apart – but with increasing $g$ they gradually coalesce at the exceptional point (EP) where $g = 2\kappa - \gamma_2 + \gamma_1$. In contrast, the imaginary components of the eigenvalues stay together before the EP and only bifurcate afterwards. If we choose a value of $\gamma_2$ higher than $\kappa$, $\gamma_2 = 7$, the location of the EP now shifts to the positive imaginary half. In Fig. S4c, the initial separation between Re{$\lambda$} is now smaller whereas in Fig. S4d, the final distance between Im{$\lambda$} is now much larger compared to the previous case. Indeed, Im{$\lambda$} crossing the zero value and switching sign (the green circle) heralds the onset of lasing which takes place in the unbroken symmetry mode (U) in Fig. S4a-b while in Fig. S4c-d it instead occurs in the broken symmetry mode (B).

Interestingly, the transition point between these behaviors is still the same value of gain as the EP, i.e. $g = 2\kappa + \gamma_1 - \gamma_2$, when detuning is introduced (Fig. S5). We conclude that, even in the presence of detuning, lasing can be either initiated in an unbroken PT-like mode if the loss in the attenuating component of the coupled cavity configuration is less than the effective coupling, i.e. $\gamma_2 < \kappa$; or in an a broken PT-like mode for the case $\gamma_2 > \kappa$. More importantly, the two PT-symmetric phases have a physical significance in the following sense: Unbroken symmetry implies that the fields in both components of the system behave in a similar fashion, i.e. with increasing gain (loss), $|a|^2$ and $|b|^2$ increase (decrease) in synchrony. However, in the broken symmetry domain, a disparity exists between the behaviors of the fields in the two components. Specifically, the ratio between the field intensity in the amplifying and lossy components $|a|^2/|b|^2$ starts to increase with gain. The bifurcation in the imaginary parts of the eigenvalues after the EP indicates this, as shown in Fig. S5.

The scope of the linear analysis presented in this section is not just limited to determining the lasing conditions, but also has important ramifications on the full nonlinear response of the system. The two distinct regimes of unbroken and broken PT-symmetry still manifest themselves in the presence of random frequency detuning and gain saturation. Remarkably, the transition between the two still arises at the point where the loss in the attenuating cavity equals the coupling strength $\gamma_2 = \kappa$. This behavior is depicted in Fig. 5 of the main text where instead of gain, the loss $\gamma_2$ is gradually increased from a low to a very high value including a passage through the point $\gamma_2 = \kappa$. In affirmation of this statement, the intensity emitted from the amplifying sub-cavity displays a dip at this point.

## S7. Probability distribution for detuning

The term that describes the resonance detuning between the coupled subcavities (that is, $\Delta$ in Eq. S3.4 and Eq. S3.5) occurs because of random perturbations in the fibers that have their origin in thermal or mechanical fluctuations. Since the fibers are relatively long (> 6 m), length expansions or contractions can easily lead to changes in the path length on the order of a micrometer, which in turn can cause a phase accumulation of $2\pi$. This would result in a frequency-detuning equivalent to half the free spectral range ($\omega_{FSR}$) of a single fiber subcavity, as shown in Fig. S6.

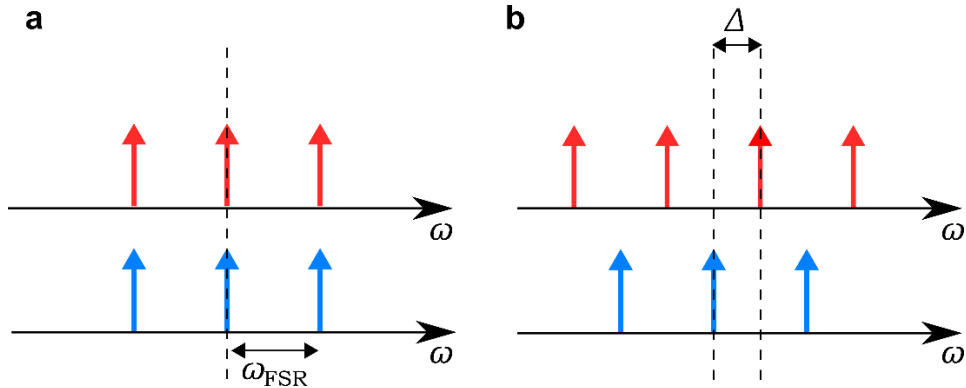

**Figure S6 | Resonances of the sub-cavities in absence and presence of detuning. a**, The resonances of the two sub-cavities are aligned identically in absence of detuning. The cavity free spectral range is $\omega_{FSR}$. **b**, In presence of detuning $\Delta$, the resonance frequencies of the two coupled sub-cavities do not align.

In principle, $\Delta$ can span the entire range $-\omega_{FSR}/2 \leq \Delta \leq \omega_{FSR}/2$. Two possible probability distributions for $\Delta$ are the uniform (Fig. S7a) and Gaussian (Fig. S7b) distributions. Because the detuning-phase-accumulation spans the range $(-\infty, \infty)$ in the Gaussian distribution, we fold the probability distribution function after each period of $\pi$ equivalent to $\omega_{FSR}/2$.

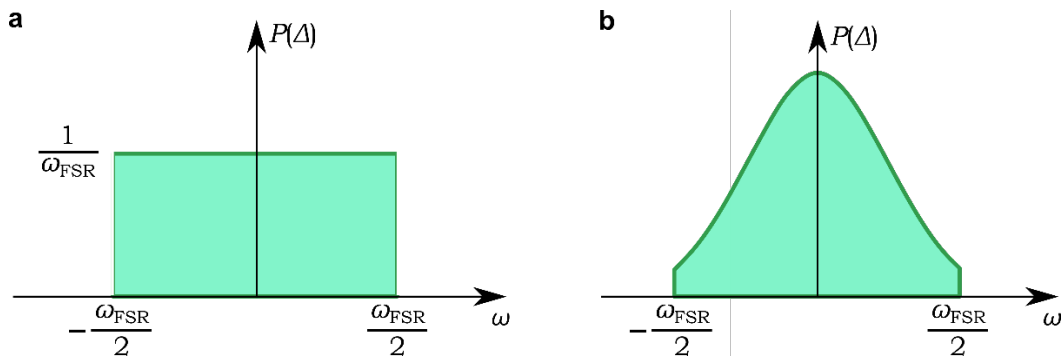

**Figure S7 | Probability distributions $P(\Delta)$ for the detuning $\Delta$ between the two coupled sub-cavities. a**, A uniform probability distribution of width $\omega_{FSR}$. **b**, A folded Gaussian probability distribution .

The Gaussian probability distribution function is $P(\Delta) = 1/(\sigma\sqrt{2\pi})\exp(-\Delta^2/2\sigma^2)$, where the full width half maximum (FWHM) is given by $2\sigma\sqrt{2\ln 2}$. To obtain solutions of the system of Eqs. (2-3) of the main text, we take an ensemble average of the steady state intensities over all the values of $\Delta$, taking into account the relevant probability distribution,

$$\langle I_{a,b}^{(ss)}\rangle = \int_{-\omega_{\text{FSR}}/2}^{\omega_{\text{FSR}}/2} I_{a,b}^{(ss)}(\Delta)P(\Delta)d\Delta, \tag{S7.1}$$

In Eq. S7.1, $I_{a,b}^{(ss)}(\Delta)$ is the stead state ('ss') intensity value obtained from a Runge-Kutta simulation for a given set of parameters $(\gamma_1, \gamma_2, g, \kappa, \Delta)$, i.e. while $\Delta$ is considered deterministic. It is only after we obtain $I_{a,b}^{(ss)}(\Delta)$ for all values of $\Delta$, that the averaging is carried out according to Eq. S7.1. For computational convenience, this integral can also be approximated by a summation after considering a finite number of sampling points for $\Delta$ in the interval $[-\omega_{\text{FSR}}/2, \omega_{\text{FSR}}/2]$. Upon analyzing the numerical results and the experimental data, we found the Gaussian distribution to be a significantly better match when compared with the uniform distribution. Moreover, the standard deviation was also found to be small, i.e. $\sigma \sim 0.1\omega_{FSR}$. The main text shows results that correspond only to the Gaussian distribution, e.g. Fig. 5. A figure similar to that when one considers a uniform distribution for $\Delta$, is shown in Fig. S8. Qualitatively, the uniform distribution leads to a larger split in the gain and loss cavity intensities since the two cavities are in essence decoupled from each other most of the time. On the other hand, a phase transition around the exceptional point more clearly visible in the case of a narrow Gaussian distribution as shown in Fig. 5 of the main text.

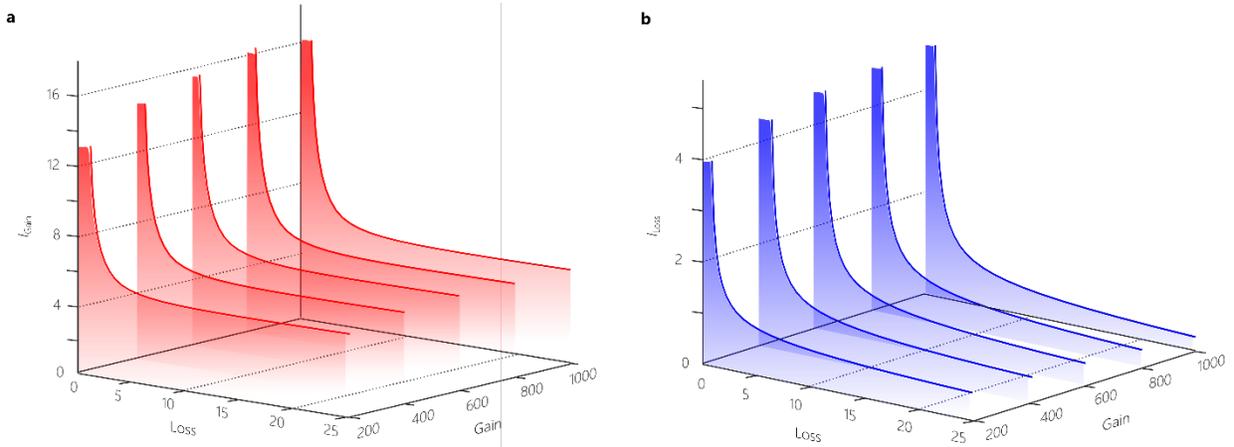

**Figure S8 | Output lasing powers assuming a uniform distribution for the detuning Δ. a,** Simulations for $I_{\text{Gain}}$ at different values of $G$ obtained from Eqs. (2-3) of the main text, based on a uniform probability distribution for $\Delta$. **b,** Same as (**a**) but for power from the loss port $I_{\text{Loss}}$.

**References**

[S1] P. Yeh and A. Yariv, *Photonics: Optical Electronics in Modern Communication* (Oxford Uni. Press, 2007), Chap. 6.

[S2] S. Longhi, "PT-symmetric laser absorber," Phys. Rev. A **82**, 031801(R) (2010).

[S3] A. K. Jahromi and A. F. Abouraddy, "Observation of Poynting's vector reversal in an active photonic cavity," Optica **3**, in press (2016).

[S4] B. E. Little, S. T. Chu, H. A. Haus, J. Foresi, and J.-P. Laine, "Microring resonator channel dropping filters," J. Lightwave Technol. **15**, 998 (1997).

[S5] M. B. Spencer and W. E. Lamb, "Theory of two coupled lasers," Phys. Rev. A **5**, 893 (1972).

[S6] M. B. Spencer and W. E. Lamb, "Laser with a transmitting window," Phys. Rev. A **5**, 884 (1972).

[S7] T. Kato, *Perturbation theory for linear operators* (Springer, Berlin, 1966).

[S8] W. D. Heiss, "Repulsion of resonance states and exceptional points," Phys. Rev. E **61** 929 (2000).

[S9] A. Guo, G. J. Salamo, D. Duchesne, R. Morandotti, M. Volatier-Ravat, V. Aimez, G. A. Siviloglou, and D. N. Christodoulides, "Observation of PT-symmetry breaking in complex optical potentials," Phys. Rev. Lett. **103**, 093902 (2009).